%% file: main.tex
\pdfoutput=1
\documentclass[a4paper,11pt]{article}
\usepackage[tableposition=above,font=small,labelfont={small,bf},labelsep=period]{caption}
\usepackage{jcappub_mod}

\usepackage[T1]{fontenc}
\usepackage{lmodern}

\usepackage{bigdelim}
\usepackage{booktabs}
\usepackage[noabbrev]{cleveref}
\crefname{equation}{eq.}{eqs.}
\Crefname{equation}{Equation}{Equations}
\crefname{enumi}{}{}
\Crefname{enumi}{}{}
\usepackage{colortbl}
\usepackage{enumitem}
\usepackage{fontawesome5}
\usepackage{longtable}
\usepackage{makecell}
\usepackage{multirow}
\usepackage{siunitx}
\DeclareSIUnit{\Mpc}{Mpc}
\DeclareSIUnit{\ueV}{\micro\eV}
\DeclareSIUnit{\elcharge}{\textit{e}}
\usepackage{slashed}
\usepackage{xspace}

\newcommand{\updated}[1]{#1}%{{\color{red} #1}}

\input{macros}

\input{journal_macros}
\begin{document}

\title{\huge Catalogues of Cosmologically Self-Consistent Hadronic QCD Axion Models}

\date{\today}

\author[a]{Luca Di~Luzio,}
\author[a,b]{Sebastian Hoof,}
\author[c,d]{Coenraad Marinissen,}
\author[\,c,d]{and Vaisakh Plakkot}

\affiliation[a]{Istituto Nazionale di Fisica Nucleare -- Sezione di Padova,\\
Via F.\ Marzolo 8, 35131 Padova, Italy}
\affiliation[b]{Dipartimento di Fisica e Astronomia ``Galileo Galilei,''
{Universit\`a} degli Studi di Padova,\\ Via F.\ Marzolo 8, 35131 Padova, Italy}
\affiliation[c]{Institute of Physics, University of Amsterdam,\\ Science Park 904, 1098 XH Amsterdam, The Netherlands}
\affiliation[d]{Theory Group, Nikhef, \\Science Park 105, 1098 XG Amsterdam, The Netherlands}
\emailAdd{luca.diluzio@pd.infn.it}
\emailAdd{hoof@pd.infn.it}
\emailAdd{c.b.marinissen@uva.nl}
\emailAdd{v.plakkot@uva.nl}

\abstract{
We extend the catalogue of ``phenomenologically preferred'' hadronic axion models to include heavy fermion representations associated with higher-dimensional decay operators.
The latter have recently been shown to self-consistently trigger a period of early matter domination, making the underlying axion models cosmologically viable.
After identifying all possible representations up to decay operator dimension $d \leq 9$, we update the hadronic axion band for the axion-photon coupling.
\updated{The central regions of the axion band are similar to those found previously and approximately independent of the axion decay constant $f_a$, suggesting that they are robust predictions and targets for future axion searches.
Moreover, we find that} $d = 6$ and $d = 7$ operators \updated{can} lead to two new viable ``model islands'' around $f_a \sim 10^{12}$\,GeV and $f_a \sim 10^{14}$\,GeV, i.e., beyond the standard post-inflationary mass region.
}

\maketitle
%\tableofcontents

\section{Introduction}\label{sec:intro}

The QCD~axion~\cite{Weinberg:1977ma,Wilczek:1977pj} is a hypothetical particle, which addresses the strong $CP$ problem~\cite{Peccei:1977hh,Peccei:1977ur} -- the observed smallness of the electric dipole moment of the neutron, $|d_n| < \SI{1.8e-13}{\elcharge\femto\m}$ (90\% CL) \cite{2001.11966} -- and provides an excellent dark matter~(DM) candidate~\cite{Preskill:1982cy,Abbott:1982af,Dine:1982ah,Turner:1983he,Turner:1985si}.
The vast landscape of axion models~\cite{2003.01100}, including KSVZ (hadronic)~\cite{Kim:1979if,Shifman:1979if} and DFSZ~\cite{Zhitnitsky:1980tq,Dine:1981rt} models, are currently being targeted by a wide range of experimental searches~\cite[e.g.][]{1801.08127}.
The main challenge for finding QCD~axions is that they are associated with an unknown energy scale parameter, the axion decay constant \fax, which could span many orders of magnitude.
Moreover, QCD~axion couplings to the Standard Model~(SM) depend in most cases on the model's ultraviolet (UV) completion, making it necessary to map out the theory space to provide search targets for experimental searches.
To this end, phenomenological selection criteria have been proposed to single out \pref models~\cite{1610.07593,1705.05370}, and catalogues of DFSZ and KSVZ models have recently become available~\cite{2107.12378,2302.04667}.

The selection criteria mentioned above are especially powerful for KSVZ models, where avoiding Landau poles (LPs) at low-energy scales renders the number of models finite, making it possible to create a complete catalogue~\cite{2107.12378}.
Another selection criterion requires that the heavy, coloured fermions \newQ in KSVZ models decay fast enough to avoid the stringent limits on metastable coloured particles, limiting the dimension of the possible decay operators in standard cosmology to $d \leq 5$~\cite{1610.07593}.
While the introduction of a nonstandard cosmological evolution typically requires the addition of new physics, it was recently pointed out that the \newQ particles can self-consistently trigger a phase of early matter domination~(EMD), making decay operators with $d > 5$ and their associated KSVZ models viable~\cite{2310.16087}.

The purpose of this paper is to revisit the catalogue of hadronic axion models in light of the findings presented in ref.~\cite{2310.16087}.
After reviewing hadronic QCD~axion models and selection criteria in \cref{sec:ksvz_models}, we first determine all possible decay operators up to $d \leq 9$ and their associated \newQ representation in \cref{sec:methods}.
We then identify those representations that can only come from higher-dimensional operators, i.e., are not compatible with any $d \leq 5$ operator, and extend the existing KSVZ catalogues by considering cosmological constraints.
We then present our results and discuss the implications for axion searches in \cref{sec:results} before concluding in \cref{sec:conclusions}.

\section{Hadronic QCD axion model building and selection criteria}\label{sec:ksvz_models}

We extensively reviewed KSVZ model building in a previous study~\cite[Sec.~II]{2107.12378} and thus only provide a summary below.
Let us denote the SM representation of a particle by $(\mathcal{C},\mathcal{I},\mathcal{Y})$, where $\mathcal{C}$, $\mathcal{I}$, and $\mathcal{Y}$ are respectively the $\gr{SU}(3)_\mathcal{C}$ colour, $\gr{SU}(2)_\mathcal{I}$ isospin, and $\gr{U}(1)_\mathcal{Y}$ hypercharge representations.

\subsection{Hadronic QCD axion theory}

The Lagrangian for hadronic axion models reads
\begin{equation}
    \lagr = \ii\,\bar{\newQ} \slashed{D} \newQ - (y_\newQ \bar{\newQ}_L \newQ_R \Phi + \text{h.c.}) - \lambda_\Phi \left( |\Phi|^2 - \frac{v_a^2}{2} \right)^2 + \dots , \label{eq:lagr} 
\end{equation}
where $y_\newQ$ is the Yukawa coupling, the Peccei--Quinn~(PQ) charge -- i.e.\ the charge corresponding to the global anomalous \UPQ PQ symmetry -- of the complex scalar field $\Phi \sim (1,1,0)$ is normalised to charge $\mathcal{X}_\Phi = 1$, and all SM fields are uncharged under the PQ symmetry.
The field $\Phi$ in the PQ-broken phase gives rise to the axion as the angular degree of freedom, and the heavy, quark-like fermion \newQ obtains a mass $\mQ = y_\newQ v_a/\sqrt{2}\sim f_a$.\footnote{Although in principle $\fax = v_a/(2N)$, we will assume $\sqrt{2}y_\newQ N$ to be an $\mathcal{O}(1)$ parameter, which might not be true for small couplings or large $N$.}

Given a representation under which \newQ transforms with PQ charge $\mathcal{X} = \mathcal{X}_L - \mathcal{X}_R = \pm 1$,\footnote{The choice $\mathcal{X} = -1$ requires the replacement of $\Phi$ with its complex conjugate, $\Phi \mapsto \Phi^*$, in \cref{eq:lagr}.} the electromagnetic and colour anomalies can be calculated as
\begin{equation}
    E = \mathcal{X} \, d(\mathcal{C}) \, d(\mathcal{I}) \left(\frac{d(\mathcal{I})^2-1}{12}+\mathcal{Y}^2\right) \quad \text{and} \quad N = \mathcal{X} \, d(\mathcal{I}) \, \dynk(\mathcal{C}) , \label{E_and_N}
\end{equation}
where $d(\cdot)$ denotes the representation's dimension and $\dynk(\mathcal{C})$ is the $\gr{SU}(3)_\mathcal{C}$ Dynkin index~(see ref.~\cite{Slansky:1981yr}).
For a model with multiple \newQs, which respectively have anomaly coefficients $E_i$ and $N_i$, the overall anomaly ratio is simply
\begin{equation}
    \frac{E}{N} = \frac{\sum_i E_i}{\sum_i N_i} \, . \label{eq:E_over_N}
\end{equation}
The axion-photon coupling \gagg then depends on $E/N$, in addition to a model-independent contribution from axion-meson mixing,
\begin{equation}
    \gagg = \frac{\alphaEM}{2\pi\fax} \, \cagg = \frac{\alphaEM}{2\pi\fax} \left(\frac{E}{N} - \caggzero\right) , \label{eq:gagg}
\end{equation}
where $\alphaEM \approx 1/137$ is the fine-structure constant and $\caggzero = \num{1.92(4)}$~\cite{1511.02867}.\footnote{References~\cite{2003.01625,2301.09647} respectively obtained for this quantity \num{2.05(3)} (using three active flavours instead of two) and \num{2.07(4)} (following the same strategy, but including the uncertainties from the quark mass ratios). More recently, ref.~\cite{2411.06737} computed \num{1.89(2)} by including the $\eta'$ meson.}
The QCD axion mass at low energies can be computed from chiral perturbation theory~\cite{1511.02867,1812.01008}:
\begin{equation}
    \ma = \SI{5.69(5)}{\micro\eV} \left(\frac{\SI{e12}{\GeV}}{\fax}\right) .\label{eq:axion_mass}
\end{equation}
At higher temperatures, lattice QCD has been used to compute the temperature-dependent QCD~axion mass $\ma(T)$~\cite{1512.06746,1606.03145,1606.07494}.

\subsection{Selection criteria for \pref models}\label{sec:selection_criteria}

Here we briefly summarise the selection criteria that have been proposed for \pref axion models in refs.~\cite{1610.07593,1705.05370,2107.12378} and discuss possible extensions.

\begin{enumerate}[label=(C\arabic*)]
    \item \newQ lifetime and decay constraints. Heavy fermions with $\mQ \lesssim \SI{e3}{\GeV}$ and electromagnetic or strong interactions are disfavoured by various searches at the LHC.
    For heavier masses, the freeze-out abundance of the \newQs typically exceeds the observed amount of DM, leading to the requirement that they must be unstable and decay~\cite{1610.07593,1705.05370}.
    However, any such decays should happen before the onset of Big Bang Nucleosynthesis (BBN) not to cause discrepancies in the observed abundance of light elements.\footnote{In this work, we will compute the effective equation of state of the Universe around the BBN temperature, cf.\ \cref{eq:eos}, to ensure that the cosmology is sufficiently radiation-dominated. More rigorously, one should compute the abundance of light elements, e.g.\ using codes such as \code{AlterBBN}~\cite{1106.1363,1806.11095} and \code{ACROPOLIS}~\cite{2011.06518}, and compare to the observed data.}
    \label{criterion:lifetime} 
    \item Axion DM overproduction. The amount of axion DM has to respect the cold DM limit of $\OmegaDM < 0.12$~\cite{1807.06209}, leading to an upper bound on $\fax \sim \mQ$.\footnote{The initial conditions of the realignment mechanism depend on the PQ~symmetry breaking scenario, which we discuss in more detail in \cref{sec:realignment}. The contribution from topological defects, i.e.\ cosmic strings and domain walls~(DWs), in the post-inflationary PQ breaking scenario can be significant~\cite{1806.04677,1906.00967,2007.04990,2108.05368}. However, these results extrapolate the outcome of numerical simulations by more than 60~orders of magnitude, and are hence subject to potentially large systematic uncertainties. Still, the limit on \fax obtained when neglecting this contribution is conservative since including it would make the allowed \fax region smaller.}\label{criterion:dm}
    \item Strong $CP$ problem. The \newQs can have opposite PQ charges with $N = \sum_i N_i = 0$. Such models do not solve the strong $CP$ problem, and we require $N \neq 0$.
    \item Landau poles. The presence of (multiple) ``large'' \newQ representations can lead to Landau poles (LPs) for the gauge couplings in the UV. This is undesirable up to the Planck scale $\MP = \SI{1.221e19}{\GeV}$, where quantum gravity effects are expected to appear. The LP criterion, which is the single most powerful criterion in the context of this work, thus requires that LPs do not arise at energies below a threshold $\LPthresh \lesssim \MP$~\cite{1610.07593,1705.05370}.\label{criterion:lp}
\end{enumerate}

We note that the lifetime constraints~\cref{criterion:lifetime} are 
not required in the pre-inflationary PQ~symmetry breaking scenario because the \newQs can get diluted by inflation.
Moreover, the energy injection due to decaying particles can lead to spectral distortions of the CMB which, however, do not seem to apply to \mQ values that we consider (see, e.g., ref.~\cite{1910.04619}).

Further constraints on the axion EFT from perturbative unitarity are derived in \cref{sec:unitarity}.
However, these turn out to only apply to large values of $E/N \gtrsim 1000$, and do not play a role in limiting the number of models in catalogues when imposing the LP criterion~\cref{criterion:lp}.

In addition to the selection criteria above, which are theoretical or observational limitations, one may also consider other desirable properties that hadronic axion models should fulfil.
These \textit{desiderata} include:
\begin{enumerate}[label=(D\arabic*)]
    \item Domain walls. Models with $\NDW \equiv 2N = 1$ do not suffer from the ``domain wall problem''~\cite{Zeldovich:1974uw}, where the energy density of domain walls dominates the energy budget of the Universe, contradicting observations~\cite{Sikivie:1982qv,Vilenkin:1982ks,Barr:1986hs}. However, other solutions to this problem exist, meaning that this property is not strictly required~\cite[e.g.][]{Sikivie:1982qv,Kim:1986ax,2003.01100}.\label{des:dw}
    \item Axions as DM. While axions may only constitute a fraction of the observed DM density, thus respecting the criterion~\cref{criterion:dm}, a single explanation for DM would be highly economical. However, the uncertain additional contributions from decaying topological defects and the uncertainties about the initial conditions for the realignment mechanism makes it difficult to apply this in practice.
\end{enumerate}

\section{Methodology for updating the model catalogues}\label{sec:methods}

We extend the catalogue of hadronic axion models~\cite{2107.12378,Zenodo_KSVZCatalogue} in two steps.
First, we identify all \newQ representations related to $d \leq 9$ operators that respect the LP criterion \cref{criterion:lp} with $\LPthresh = \SI{e18}{\GeV}$ (see \cref{sec:new_reps}).
We are in particular interested in representations that cannot arise from lower-dimensional operators.
This is because lower-dimensional operators determine the cosmological evolution of the \emph{individual} \newQs.
We then assess the cosmological evolution of models with multiple \newQs in a second step in \cref{sec:cosmo}.

\subsection{New representations for higher-dimensional operators}\label{sec:new_reps}

We use the software code \deco~\cite{2212.04395} (version tag v1.1) to create a nonminimal but exhaustive list of operators for a given mass dimension and field content, and their charges, under specified symmetries. 
From these inputs, \deco computes the nonredundant operators at the specified mass dimension.
Here, the field content consists of the SM fields and the heavy fermions \newQ, and we employ the SM gauge symmetries.
Note that the SM charges of the \newQs are \textit{a priori} unknown, meaning that a scan over all possible charges is required.
The PQ scalar field is not included at this point as a simplification, since it does not affect the SM charges of the \newQs.
Further details on the workflow of \deco are described in \cref{sec:deco}.

For higher operator dimensions, it becomes computationally very expensive to scan over all representations that may induce $\newQ \to \text{SM}$ decay operators, and we use the restrictive LP criterion~\cref{criterion:lp} as a preselection.
To compute the LP, we evolve the two-loop renormalisation group equations for the running of the couplings, following ref.~\cite{2107.12378} (see also refs.~\cite{Machacek:1983tz,1504.00359}), using $\alpha_1(M_Z) = 0.01692$, $\alpha_2(M_Z) = 0.033735$, and $\alpha_3(M_Z) = 0.1173$~\cite{1208.3357} at $M_Z = \SI{91.188}{\GeV}$~\cite{ParticleDataGroup:2024cfk}, neglecting the associated uncertainties.\footnote{Compared to a previous study~\cite{2107.12378}, we now employ slightly refined settings when computing the LPs and, most importantly, a root finding algorithm which more precisely estimates the LPs, i.e.\ the first zero of any of the gauge couplings $\alpha_i^{-1}$. This leads to sub-percent-level differences in the computed LPs and small differences in the size of the computed catalogues, which do not impact the main conclusions of this work.}

\newQs with ``large'' SM charges induce LPs at lower energies, which allows us to restrict the allowed charges of \newQ even before computing the entire set of operators using \deco.
We provide analytical estimates for the resulting bounds at the one-loop level in \cref{sec:analyticalapprox}, which approximately agree with the two-loop beta functions that we use in this work.
With this, the ``edges'' of the SM charges for \newQ were identified for each gauge group, setting $\mQ = \SI{e17}{\GeV}$ and one fermion, $\NQ = 1$.
These choices are conservative since larger charges are incompatible with the LP criterion and can safely be ignored.

We then iterate over all possible combinations of the \newQs' SM charges within the allowed ``edges'' and save the results (the list of operators) whenever the \newQ~representation allows for $\newQ \to \text{SM}$ decay operators.
This process was repeated for all mass dimensions $3 \leq d \leq 9$.
We provide a few examples for new operators and their associated representations in \cref{tab:some_reps}, while the complete list of all representations for $3 \leq d \leq 9$ can be found in in \cref{tab:all_reps}.
In particular, for $3 \leq d \leq 5$, we find the same representations as refs.~\cite{1610.07593,1705.05370}.

\Cref{tab:some_reps} lists the anomaly ratios, DW number, and example operators for a given \newQ representation, in addition to the lowest mass dimension at which the representation appears and the LP scale for $\mQ = \SI{e17}{\GeV}$.
We stress that the example operators given in \cref{tab:some_reps,tab:all_reps}, and the complete list in the supplementary material~\cite{Zenodo_UpdatedKSVZCatalogue}, do not necessarily provide a nonredundant basis of effective operators.
This is not an issue for the scope of this study, or previous works, since the \emph{possible existence} of a decay operator is sufficient to make the associated representation a viable candidate for axion models.

\begin{table}
    \centering
    \caption{Examples of LP-allowed \newQ representations for $\NQ = 1$ and $\mQ = \SI{e17}{\GeV}$. The columns give the three SM charges, the resulting anomaly ratio and domain wall number, the lowest mass dimension where the representation can appear, an example operator, and the value of the first Landau pole. The complete list for $d \leq 9$, and a more detailed description, is provided in \cref{tab:all_reps}.\label{tab:some_reps}}
    \begin{tabular}{rS[table-format=2]S[table-format=-2]ccccS[table-format=1.1e2]}
    \toprule
    \multicolumn{3}{c}{Rep.\ $(\mathcal{C},\mathcal{I},6\mathcal{Y})$} & $E/N$ & $\NDW$ & Minimal $d$ & Example operator & \multicolumn{1}{c}{LP [GeV]} \\
    \midrule
     3 & 1 & -14 & 98/3 & 1 & 6 & $\bar{\newQ}_L\,d_R\,(\bar{e}_R^c\,e_R)$ & 2.2e+22 \\
     $\overline{3}$ & 1 & 8 & 32/3 & 1 & 6 & $\bar{u}_R\,\gamma_\mu\,e_R\,\bar{d}_R\,\gamma^\mu\,\newQ_R$ & 3.0e+28 \\
     $\overline{3}$ & 1 & -10 & 50/3 & 1 & 6 & $(\bar{d}_R\,d_R^c)\,\bar{e}_R\,\newQ_L$ & 6.4e+25 \\
     \multicolumn{8}{c}{\vdots} \\
     \midrule
     3 & 2 & -17 & 149/3 & 2 & 7 & $H^\dagger\,\bar{\newQ}_L\,d_R\,(\bar{e}_R^c\,e_R)$ & 1.2e+19 \\
     3 & 2 & 19 & 185/3 & 2 & 7 & $H\,\bar{\newQ}_L\,u_R\,(\bar{e}_R\,e_R^c)$ & 4.4e+18 \\
     3 & 3 & -14 & 110/3 & 3 & 7 & $H^\dagger\,(\bar{u}_R\,u_R^c)\,\bar{\newQ}_R\,\ell_L$ & 1.2e+19 \\
     \multicolumn{8}{c}{\vdots} \\
     \bottomrule
    \end{tabular}
\end{table}

Before investigating the cosmological consequences of the new representations, let us summarise a number of interesting points related to the extended list of relevant representations.
Overall, for $\NQ = 1$, we find 43, 44, 20, and 14 representations that respect the LP criterion~\cref{criterion:lp} for $d = 6$ to $d = 9$ operators, respectively, at the conservative value of $\mQ = \SI{e17}{\GeV}$.
While the operators with $d = 8$ and $d = 9$ will turn out to be cosmologically irrelevant in the post-inflationary scenario, we include them for completeness.
For lower values of \mQ, the number of new representations will be lower and, as we will see, updating the catalogues becomes combinatorially tractable.
This is also aided by the fact that, for $\NQ > 1$, we can only add a few \newQs with the new, ``large'' representations for $d > 5$ before hitting a LP below \LPthresh.

We further note the possibility to have a model with a single $\newQ \sim (8,1,0)$ for $d = 6$, or $\newQ \sim (27,1,0)$ or $(35,1,0)$ for $d = 8$, corresponding to $E/N = 0$.
This is the anomaly ratio of the original KSVZ model, which was actually not a \pref model in the classification of ref.~\cite{1705.05370}.
Since $E/N = 0$ is a widely used benchmark model, it is reassuring to have a \pref model with $\NQ = 1$, whereas this was only possible for $\NQ > 1$ in the previous catalogue~\cite{Zenodo_KSVZCatalogue}.

Moreover, since many of the new representations transform in the fundamental of $\gr{SU}(3)_\mathcal{C}$ and trivially under $\gr{SU}(2)_\mathcal{I}$, we can anticipate new models with $\NDW = 1$ that hence avoid the DW problem, cf.\ desirable feature~\cref{des:dw}.

\subsection{Cosmological evolution of the new models}\label{sec:cosmo}

To solve the cosmological evolution in the presence of the \newQs, we write a Boltzmann equation solver in Python (see \cref{sec:emd}).
The resulting evolution of the Hubble parameter~$H$, including potentially a phase of EMD, is then used to compute the axion realignment density in \cref{sec:realignment}.

Since alternative cosmological histories and the related Boltzmann equations have been studied in detail in the past, we only summarise the equations used and some technical aspects for those who wish to reproduce our results.

\subsubsection{Early matter domination}\label{sec:emd}

Early matter domination and other nonstandard cosmologies (see ref.~\cite{2006.16182} for a recent review) have been extensively studied in the context of axion physics~\cite{hep-ph/0005123,0711.1352,0912.0015,1808.01879,1904.05707,1911.07853,2104.03982,2107.13575,2107.13588,2110.12253,2308.01352,2310.16087}.
A phase of EMD is often realised in a low-temperature reheating scenario but, as the authors of ref.~\cite{2310.16087} pointed out, hadronic axion models can self-consistently induce a phase of EMD due to the presence of the \newQs.
Since we want to extend previous results to consider multiple \newQs, we cannot rely on a simple parametrisation of the EMD scenario and instead adapt the relevant Boltzmann equations~\cite{1906.04183,2310.16087} to compute the alternative cosmologies numerically:
\begin{align}
     \frac{\dd T}{\dd u} &= \left(1 + \frac{T}{3\gS}\frac{\dd \gS}{\dd T}\right)^{-1} \left[-T + \sum_i \frac{\Gamma_i \rho_i}{3 H s} \right] , \label{eq:boltzmann:te} \\
    \frac{\dd n_i}{\dd u} &= -3 \, n_i - \frac{\Gamma_i}{H} \left(n_i - n_{i,\text{eq}}\right) - \frac{\sigv}{H} \left(n_i^2 - n_{i,\text{eq}}^2\right) , \label{eq:boltzmann:nQ}
\end{align}
where $T$ is the temperature of the heat bath, $n_i$ are the number densities of the \newQs, $u \equiv \ln(a/a_0)$ with the scale factor $a$ relative to some arbitrary scale $a_0$, \gS denotes the effective degrees of freedom (DOF) in entropy density $s$ of the thermal bath, for which we use the results of ref.~\cite{1803.01038}, and $\rho_i$ are the energy densities of the \newQs.
At temperature around or below \mQ, it is sufficient to approximate the energy of the \newQs using the average thermal temperature, $E_\newQ \approx \mQ + 3T/2$, to compute their energy density $\rho_i = E_\newQ \, n_i$.
Note that the variable $u$ is a good choice for EMD and other nonstandard cosmologies since the relation between $T$ and physical time~$t$ may not be bijective.
In \cref{eq:boltzmann:nQ}, we neglect entropy injection from annihilations and the role of inverse decays.

The decay rates $\Gamma_i$ in \cref{eq:boltzmann:nQ} for the various \newQ operators can be characterised by the operator mass dimension~$d$ and depend on the EFT suppression scale \Ldec.
In particular, we have $\Gamma^{(4)} = \mQ/8\pi$ and
\begin{equation}
    \Gamma^{(d > 4)} = \frac{\mQ}{4\,(4\pi)^{2\mtx{N}{fin} - 3}\,(\mtx{N}{fin} - 1)!\,(\mtx{N}{fin} - 2)!} \left(\frac{\mQ}{\Ldec}\right)^{2(d-4)} \label{eq:decay_ops}
\end{equation}
for $d > 4$ (see, e.g., ref.~\cite{1504.00359}), where $\mtx{N}{fin}$ is the number of final state particles.
Given the hierarchy of decay operators in \cref{eq:decay_ops}, resulting in a $(\mQ/\Ldec)^2$ suppression for $(d+1)$-dimensional compared to $d$-dimensional operators, it is sufficient to only consider the lowest-dimensional operator for each \newQ in \cref{eq:boltzmann:nQ}.
For our main results, we set $\Ldec = \MP$, but we discuss the effect of lowering \Ldec in \cref{sec:caveats}.

The thermally-averaged scattering cross section in \cref{eq:boltzmann:nQ} for \newQ is given by
\begin{equation}
    \sigv = \frac{\pi \alphaS^2}{16\mQ^2} \; \mtx{C}{ann} , \label{eq:sigv}
\end{equation}
where \alphaS is the strong coupling constant, whose running we compute at two-loop level (see \cref{sec:new_reps} for details), and $\mtx{C}{ann} \equiv c_f N_f + c_g$, where $N_f = 6$ is the number of quark flavours while $c_f$ and $c_g$ are group theory factors that parametrise the annihilation of $\newQ\bar{\newQ}$ into quarks and gluons, respectively.
In particular, we have $c_f = 2/9$ and $c_g = 220/27$ for triplets \cite{ParticleDataGroup:2016lqr}, and $c_f = 3/2$ and $c_g = 27/4$ for octets \cite{hep-ph/9806361}, such that $\mtx{C}{ann} \sim 10$ in either case. 
In what follows, we will assume the triplet case and neglect the $\mathcal{O}(1)$ variation due to the different $\mathcal{Q}$ representations.
This is also because a potentially much larger effect comes about due to potential \newQ hadronisation and bound states at temperatures below the QCD scale, $T \lesssim \mtx{T}{QCD} \approx \SI{160}{\MeV}$~\cite[e.g.][]{1812.08235}, which we elaborate on in \cref{sec:caveats}.

The initial conditions for the \newQ abundances are determined by the assumption of initial thermal equilibrium.
The corresponding number densities are given by~\cite[e.g.][]{textbook_baumann_cosmology}
\begin{align}
    n_{i,\text{eq}} &= \frac{g_i}{2\pi^2} T^3 I_+(\mQ/T) , \label{eq:n_eq} \\
    I_+(x) &= \int_0^\infty \! \dd \xi \; \frac{\xi^2}{\exp\left(\sqrt{\xi^2 + x^2}\right) + 1} = x^2 K_2(x) \simeq \left\{ \begin{array}{ll}
        3\zeta(3)/2 & \text{if $x \ll 1$} \\
        \sqrt{\pi/2} x^{3/2} \ee^{-x} & \text{if $x \gg 1$}
    \end{array} \right. ,
\end{align}
where $K_p(x)$ is the modified Bessel function of the second kind of order $p$.
For a single \newQ, we have $g_i = g_\newQ = 2$, while for multiple \newQs, we have to multiply this value by the number of identical \newQs, which we return to in \cref{sec:multipleQs}.

We compute the equilibrium number densities in \cref{eq:n_eq} at $T = 10\,\mQ$ and then proceed to solve \cref{eq:boltzmann:te,eq:boltzmann:nQ} until reaching the BBN scale temperature, which we fix at $\TBBN = \SI{1}{\MeV}$.
Since the initial temperature of $T = 10\,\mQ$ can be very large, we point out that we can only self-consistently assume a post-inflationary PQ breaking scenario if the reheating temperature is higher than this.
Its maximum value is determined by limits on the Hubble scale at the end of inflation, which can be derived in slow-roll inflation; the limits on the tensor-to-scalar ratio, $r_{0.05} < 0.036$ (95\% credible interval) at the pivot scale $k = \SI{0.05}{\Mpc^{-1}}$~\cite{2110.00483}; and the measured value of the scalar spectral amplitude $\mtx{A}{s}$, $\ln(10^{10} \mtx{A}{s}) = \num{3.047(14)}$~\cite{1807.06209}.
These imply that $\mtx{H}{I} = \sqrt{\pi \, r_{0.05}\mtx{A}{s}} \, \mP \lesssim \SI{4e13}{\GeV}$, where $\mP \equiv \MP/\sqrt{8\pi}\approx \SI{2.436e18}{\GeV}$ is the reduced Planck mass.
The related Gibbons--Hawking temperature~\cite{10.1103/PhysRevD.15.2738} $\mtx{T}{GH} = \mtx{H}{I}/2\pi \lesssim \SI{6e12}{\GeV}$ should be compared to \fax.

To render our analysis as model-independent as possible, the running of \alphaS in \cref{eq:sigv} ignores the threshold effects at \mQ, i.e., in the small window of $\mQ \lesssim T \lesssim 10\,\mQ$.
Similarly, the effective DOF (in entropy) neglect the presence of the \newQs at high temperatures.
This is justified because, above the top quark annihilation threshold of  $T \gg \SI{346}{\GeV}$, the SM has $\gRho \approx \num{107}$~\cite{1609.04979}, while we only find viable models with $\NQ < 35$.
Recall that we assume all \newQs to have identical mass, $m_i = \mQ = \fax$.

The Hubble parameter in \cref{eq:boltzmann:te,eq:boltzmann:nQ} is given by the first Friedmann equation,
\begin{equation}
    H^2(u) = \frac{1}{3\mP^2}\left(\rhoSM + \sum_i \rho_i \right) = \frac{1}{3\mP^2}\left[\frac{\pi^2}{30} \, g(T) \, T^4 + \sum_i \rho_i \right] , \label{eq:hubble}
\end{equation}
where the first term is the SM energy density with $T = T(u)$ from \cref{eq:boltzmann:te}, while we use the effective DOF in energy density $\gRho(T(u))$ as given in ref.~\cite{1803.01038}.

The consistency with BBN can be demonstrated by computing the predicted abundance of light elements and comparing it to the available cosmological and astrophysical data.
This effectively means that \newQ~decays have to happen ``before BBN'' (see \cref{sec:selection_criteria}), which we make more quantitative by computing the effective equation of state (EOS) of the Hubble parameter.
Starting from the definition $w \equiv P/\rho$, and using both Friedmann equations, $\rho = 3 \mP^2 H^2$ and $P = -2\mP^2 \dot{H} - \rho$, we can see that
\begin{align}
    w &= -\frac{2}{3 H^2} \frac{\dd H}{\dd u} \frac{\dd u}{\dd t} - 1 = -\frac{2}{3H} \frac{\dd H}{\dd u} - 1 = -\frac{1}{3H^2} \frac{\dd H^2}{\dd u} - 1 , \label{eq:eos} \\
    \frac{\dd H^2}{\dd u} &= \frac{1}{3 \mP^2} \left[\left(\frac{T}{\gRho}\frac{\dd\gRho}{\dd T} + 4\right)\frac{\dd T}{\dd u}\frac{\rhoSM}{T} + \sum_i \left(E_\newQ \, \frac{\dd n_i}{\dd u} + \frac32 \frac{\dd T}{\dd u} n_i \right) \right] , \label{eq:hubble_der}
\end{align}
where we compute $\dd \gRho/ \dd T$ analytically from its functional form given in ref.~\cite{1803.01038} and where the values of $\dd T/\dd u$ and $\dd n_i / \dd u$ are given by \cref{eq:boltzmann:te,eq:boltzmann:nQ}.
An EOS of $w = 1/3$ corresponds to a radiation-dominated Universe, while $w = 0$ indicates matter domination.
To ensure that the Universe follows standard cosmology (radiation domination around BBN), we demand that $w(\TBBN) > 0.3$.

Having all ingredients ready, we can proceed to solve \cref{eq:boltzmann:te,eq:boltzmann:nQ} using the Python method \code{integrate.solve\_ivp} from the \code{scipy} package~\cite{scipy}.
We define $u \equiv 0$ at our initial temperature, $T = 10\,\mQ$, and solve until reaching \TBBN.
Since we solve the equations over many orders of magnitude in scale factor, it is useful to focus the relative tolerance, which we set to \num{e-6}, while ignoring the absolute tolerance by setting it to~0.
We further define two ``events'' for the solver: one at the crossing point of the BBN temperature $\TBBN = \SI{1}{\MeV}$ and another when the energy density of a \newQ become drastically diluted after it decays and falls below $\rho_i < \num{e-20}\mtx{\rho}{SM}$.
Whenever the latter happens, we remove the related \cref{eq:boltzmann:nQ} from the system of equations.

\begin{figure}
    \centering
    \includegraphics[width=6in]{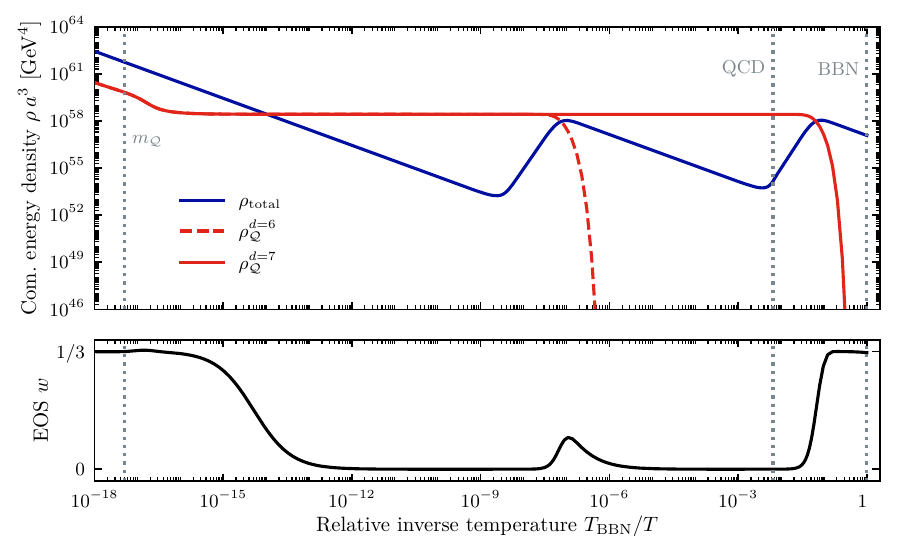}
    \caption{Example for the cosmological evolution of two \newQs with $\mQ = \SI{2e14}{\GeV}$ and associated with $d = 6$ and $d = 7$ decay operators, respectively.
    The top panel shows the evolution of their comoving energy densities (solid and dashed red lines) and the total commoving energy density.
    The bottom panel shows the effective equation of state as a function of temperature.}
    \label{fig:cosmo_evo_ex}
\end{figure}

\Cref{fig:cosmo_evo_ex} shows the solution of the Boltzmann equations for two \newQs with mass $\mQ = \SI{2e14}{\GeV}$ and lowest-dimension decay operators of $d = 6$ and $d = 7$, respectively.
First, note the period of EMD ($w = 0$) due to the \newQ freezeout.
The length of the EMD phase is determined by the \newQ with the higher-dimensional $d = 7$ decay operator, although the $d = 6$ operator causes a small perturbation from matter domination at around $T \sim \SI{e7}{\MeV} = \num{e7}\,\TBBN$.
Even small perturbations can potentially affect the axion DM abundance when they happen around the QCD crossover, $\mtx{T}{QCD} \approx \SI{160}{\MeV}$.
Around that scale, the QCD~axion starts to become dynamical, effectively determining its present-day abundance $\Omega_a h^2$, which we will compute in \cref{sec:realignment}.
In any case, the example in \cref{fig:cosmo_evo_ex} shows a period of EDM around $\mtx{T}{QCD}$ and demonstrates that, while self-consistent, the nonstandard cosmologies from the new models are not trivial.

The second observation relates to the value of the EOS at BBN, $w(\TBBN)$.
It ultimately depends on the slower, $d = 7$ decay rate, which is however fast enough to return the Universe to a radiation-dominated state before BBN, $w(\TBBN) > 0.3$.
The scenario depicted in \cref{fig:cosmo_evo_ex} thus respects criterion~\cref{criterion:lifetime}.

\subsubsection{Axion DM from realignment production}\label{sec:realignment}

Following ref.~\cite{2310.16087}, we use the \cpp code \mimes~\cite{2110.12253} (version tag v1.0.0) to compute the axion abundance today.
\mimes solves the realignment equation
\begin{equation}
    \frac{\dd^2 \theta}{\dd u^2} + \left(3 + \frac{1}{H} \frac{\dd H}{\dd u}\right) \frac{\dd \theta}{\dd u} + \left(\frac{\ma}{H}\right)^2 \sin(\theta) = 0, \label{eq:realignment}
\end{equation}
where $\theta$ is the axion field normalised by \fax, with initial conditions $\theta(0) \equiv \thetai$ and $\dot{\theta}(0) = 0$ at some sufficiently small $T(u = 0)$, for which $\ma(u = 0) \ll 3H(u = 0)$.
The evolution of the Hubble parameter $H(u)$ was computed in \cref{sec:emd}, while we use the temperature-dependent axion mass $\ma(T)$ from ref.~\cite{1606.07494}.

In the post-inflationary scenario, the Universe consists of a huge number of causally disconnected patches, each with its own value of $\thetai \sim \mathcal{U}(-\pi,\pi)$, i.e., randomly drawn from a presumably uniform distribution~\cite{Turner:1985si}.
The averaged energy density is
\begin{equation}
    \left\langle \Omega_a \right\rangle = \frac{1}{2\pi} \int_{-\pi}^{\pi} \! \dd \thetai \; \Omega_a(\thetai^2) = \frac{1}{\pi} \int_{0}^{\pi} \! \dd \thetai \; \Omega_a(\thetai^2) \equiv \Omega_a(\thetaieff) , \label{eq:postinf_avg}
\end{equation}
where, by virtue of the intermediate value theorem, the integral equals $\Omega_a$ evaluated at some effective value $\thetaieff \in (0,\pi)$.

\begin{figure}
    \centering
    \includegraphics[width=6in]{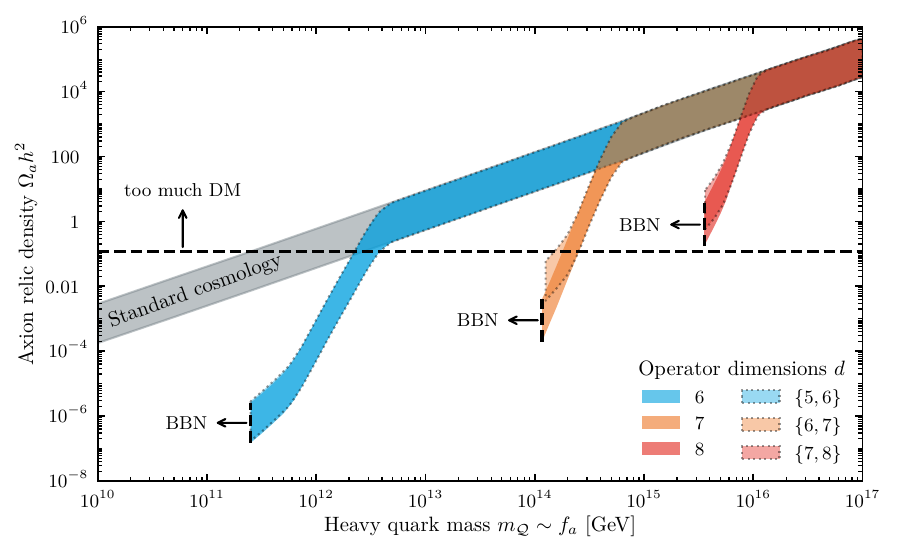}
    \caption{Axion relic density $\Omega_a h^2$ for different models. We show the range $\thetai \in [0.5,\,\pi/\sqrt{3}]$ for standard cosmology (grey shading), $\NQ = 1$ with decay operators of dimensions $d = 6, 7, 8$ (coloured shading), and $\NQ = 2$ where the second \newQ is associated with decay operators with $d = 5, 6, 7$ (coloured shading, dotted outline). The BBN bound excludes cosmologies with an EOS of $w < 0.3$ at $\TBBN = \SI{1}{\MeV}$, while the DM bound is $\OmegaDM < 0.12$~\cite{1807.06209}.}
    \label{fig:rd_d6_to_d8}
\end{figure}

In \cref{fig:rd_d6_to_d8}, we compare our Boltzmann solver to ref.~\cite[Fig.~2]{2310.16087} by showing the $\Omega_a h^2$ bands for different decay operators and values $\thetai \in [0.5,\,\pi/\sqrt{3}]$.
To highlight the possible impact of hadronic axion models with multiple \newQs, we show the nontrivial case of two \newQs with slightly different lowest-dimension decay operators associated with them, $d_1$ and $d_2$ respectively, where $d_2 = d_1 + 1$.
For $d_1 = 5$ and $d_2 = 6$, there is no visible difference between the effect of the alternative cosmology from two \newQs and the one induced by only one \newQ with $d = 6$ since the earlier decays happen well before the QCD phase transition for the allowed cosmologies with $w(\TBBN) > 0.3$.
For $d_1 = 6$ and $d_2 = 7$, the earlier decays have a stronger influence on the cosmological evolution (cf. \cref{fig:cosmo_evo_ex}), leading to a significant change in $\Omega_a h^2$ compared to the case of one \newQ. 
This is also true for $d_1 = 7$ and $d_2 = 8$ with the difference that the later decays happen closer to the onset of the axion oscillations, leading to a weaker overall effect of the alternative cosmologies and the differences between them.

As a final comment, we note that the range of \thetai is helpful for a comparison to ref.~\cite{2310.16087}.
However, the value $\thetaieff = \pi/\sqrt{3} \approx 1.81$ corresponds to averaging over a \emph{harmonic} potential.
The QCD~axion potential is not harmonic since the $\sin(\theta)$ term in \cref{eq:realignment} introduces anharmonic effects~\cite{Turner:1985si}, and thus $\thetaieff > \pi/\sqrt{3}$.
Using \mimes, we find that QCD~axions are all of the DM for $\fax \approx \SI{1.9e11}{\GeV}$ in standard cosmology, corresponding to $\thetaieff \approx 2.2$.

\section{Results}\label{sec:results}

Having extended the list of potentially viable representations and having considered the impact of multiple \newQs on the cosmological evolution, we can now proceed to construct the KSVZ model band.
We mostly follow the procedure described in ref.~\cite{2107.12378}, except that we do not fix \fax to a specific value and consider constraints from BBN to identify the viable EMD cosmologies.
Moreover, we choose $\thetai = 2.2$ to mimic the realignment production in the post-inflationary scenario.
Note that this choice neglects the contribution of topological defects to $\Omega_a h^2$, which we discuss further in \cref{sec:caveats}.

\subsection{Models with one new heavy, coloured fermion}\label{sec:oneQ}

The cosmological implications for KSVZ-type models with one \newQ can largely be deduced from the results in ref.~\cite[Fig.~2]{2310.16087}, or from \cref{fig:rd_d6_to_d8}.
This is because the cosmological evolution in \cref{eq:decay_ops} mostly depends on the dimensionality of the operators since we ignore the mild dependence of the annihilation cross section in \cref{eq:sigv} on the specific \newQ~representation (see \cref{sec:emd}).
With this approximation in place, the cosmology for $\NQ = 1$ is essentially determined by the lowest-dimensional operator, plus the value of $\mQ \sim \fax$.
The most important consequences from \cref{fig:rd_d6_to_d8} are that:
\begin{itemize}
    \item Only models with $d \leq 7$ will be cosmologically viable, corresponding to $\fax \lesssim \SI{e14}{\GeV}$, as it is otherwise not possible to sufficiently dilute the axion energy density in the post-inflationary PQ breaking scenario.\footnote{Even in absence of a concrete EMD cosmology, a similar upper bound of $\fax \lesssim \SI{e15}{\GeV}$ can be established from more generic entropy injection arguments~\cite{hep-ph/9510461}.}
    \item The operators of $d = 6$ and $d = 7$ lead to two separate, viable regions in the post-inflationary scenario around $\fax \sim \SI{e12}{\GeV}$ and $\fax \sim \SI{e14}{\GeV}$, in addition to the $\fax \lesssim \SI{2e11}{\GeV}$ region in standard cosmology.
    These intervals or ``islands'' are delimited by BBN and DM constraints for low and high \fax, respectively.
\end{itemize} 

Let us discuss the location of the islands mentioned above in more detail.
Very roughly, the \newQs start decaying when their decay rates equal the Hubble parameter, $H(\Tdec) \sim \Gamma$.
Demanding that the decays happen before BBN, $\Tdec \gtrsim \TBBN$, we find that $\mQ \gtrsim \SI{2e11}{\GeV}$ and $\mQ \gtrsim \SI{e14}{\GeV}$ for $d = 6$ and $d = 7$ operators in standard cosmology, respectively.
For $d = 5$, we have $\mQ \gtrsim \SI{e5}{\GeV}$, which is however disfavoured by hot DM constraints (see \cref{fig:ksvz_band_gagamma}), and we can safely assume that operators $d < 5$ do not induce an EMD cosmology in the parameter space of interest.

Similarly, we can estimate the location of the upper bound on \mQ from expressions for $\Omega_a h^2$ in EMD (or low-temperature reheating) cosmologies~\cite[e.g.][Eq.~(71)]{0912.0015}.
Solving the underlying equations numerically for the relevant values of $\mQ = \fax \lesssim \SI{e17}{\GeV}$, we find that the DM limit requires that $\mQ \lesssim \SI{2e12}{\GeV}$ and $\mQ \lesssim \SI{3e14}{\GeV}$ for $d = 6$ and $d = 7$ operators, respectively.

For $d = 8$, the simplistic computations above imply that $\mQ \gtrsim \SI{3e15}{\GeV}$ from the BBN constraint and $\mQ \lesssim \SI{6e15}{\GeV}$ from the DM limit.
However, when performing more careful computations, we do not find any viable models with $d = 8$, which is because the approximate solutions for $\Omega_a h^2$ in EMD are not very accurate for larger \fax and our cosmologies can deviate from pure EMD.

\subsection{Models with multiple new heavy, coloured fermions}\label{sec:multipleQs}

Since we assume that all \newQs have the same mass $\mQ = \fax$, there is only a limited number of different cosmologies that we have to consider.
In particular, the cosmology of multiple \newQs is essentially determined by the highest-dimensional operator amongst the lowest-dimensional operator for each \newQ.
Moreover, if multiple \newQs have decay operators of the same dimension, we can simply include their multiplicity in the $g_i$ factor in \cref{eq:n_eq}, and thus consider one instead of multiple separate equations for $n_i$ in \cref{eq:boltzmann:nQ}.

The different nonstandard cosmologies can thus be labelled by what we call the \textit{dimensional signature}, which we define as the multiplicities of the lowest dimension associated with the representation of each \newQ for dimensions $5 \leq d \leq 8$.
As a concrete example, consider two models with $\NQ = 3$: i) one with two \newQs with charges $(3, 3, -4/3)$ and one \newQ with charges $(3, 2, -17/6)$, and ii) another with \newQs with charges $(3,3,5/3)$, $(8,1,-1)$, and $(3,3,-7/3)$, respectively.
In both cases, two \newQs arise from $d = 5$ operators while the remaining comes from a $d = 7$ operator (see \cref{tab:all_reps}), and the corresponding dimensional signature would be $[2,0,1,0]$, i.e., two $d = 5$, no $d = 6$, one $d = 7$, and no $d = 8$ operators.
The models are thus cosmologically equivalent with our approximations.
We also saw examples of the cosmological evolution for the $[0,1,1,0]$ case in \cref{fig:cosmo_evo_ex}, while the impact on the axion DM constraints for the cases $[1,1,0,0]$, $[0,1,1,0]$, and $[0,0,1,1]$ is included in \cref{fig:rd_d6_to_d8}.

For a more systematic survey, we use root finding to determine the boundaries of the ``islands'' for $\NQ > 1$, which we discussed for $\NQ = 1$ in \cref{sec:oneQ}.
We find again two islands characterised by $d = 6$ and $d = 7$, while no models with $d \geq 8$ are viable in this scenario.
More concretely, the islands respectively span regions of $\fax \in [\num{0.25},\,\num{2.0}] \times \SI{e12}{\GeV}$ and $\fax \in [\num{1.0},\,\num{1.7}] \times \SI{e14}{\GeV}$.
We then fix a number of values for \fax within these regions and for the standard region, which we delimit by cold~\cite{1807.06209} and hot DM~\cite{2212.11926,2310.08169} constraints from cosmology, and compute the KSVZ catalogues for each \fax value.

\updated{In this context, we emphasise that \NQ is \emph{not} a free parameter of the model catalogue.
Instead, we start with $\NQ = 1$ and increase $\NQ \mapsto \NQ + 1$ until the LP~criterion~\cref{criterion:lp} is violated for all possible combinations of representations.
Once a model encounters a LP below \LPthresh, adding another \newQ would also cause a LP below \LPthresh.
The LP~criterion thus naturally sets the maximal value of~\NQ that we need to explore.}

\begin{figure}
    \centering
    \includegraphics[width=6in]{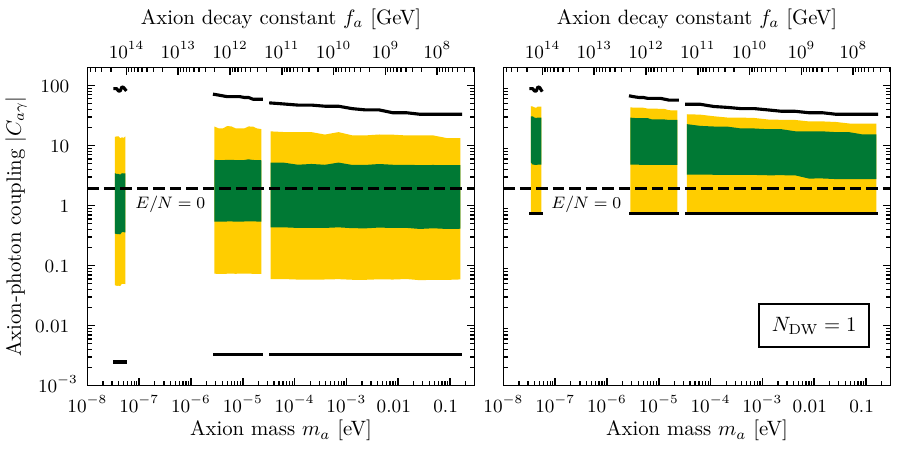}
    \caption{Self-consistent 68\% (green) and 95\% (yellow) central regions of the hadronic axion band in the post-inflationary PQ symmetry breaking scenario for $|\cagg|$. We show the band as obtained from the full catalogues (\textit{left}), and for models with $\NDW = 1$ (\textit{right}). The dashed line marks the original KSVZ model ($E/N = 0$), while the solid lines respectively delimit the minimal and maximal value of $|\cagg| = |E/N - 1.92|$, i.e., ignoring the uncertainties of \caggzero.}
    \label{fig:ksvz_band_cag}
\end{figure}
The left panel in \cref{fig:ksvz_band_cag} shows the resulting hadronic axion model band, where we used the \ma-\fax relation from \cref{eq:axion_mass} without uncertainties, but include the uncertainty on the model-independent contribution on the axion-photon coupling \gagg in \cref{eq:gagg}. 
The bands in the figure are the central 68\% and 95\% regions of the distribution from \num{2e8} Monte Carlo simulations of equally-weighted $E/N$ values of the full catalogues and \caggzero values drawn from a Gaussian distribution, $\caggzero \sim \mathcal{N}(\mu = 1.92,\, \sigma = 0.04)$.

\updated{In general, we expect the central model bands (green and yellow regions in \cref{fig:ksvz_band_cag}) to be smooth and the minimal/maximal value of $|\cagg|$ (black lines in \cref{fig:ksvz_band_cag}) to monotonically decrease/increase as \fax increases.
However, the bands exhibit subtle wiggles towards lower \fax values in the standard region and the islands.
This small effect is due to sampling noise, since the number of models can be greater than our sizeable sample size.
More interestingly, the maximal $|\cagg|$ value in the $d = 7$ island features a dip, which is not a sampling artefact.
For $\fax \gtrsim \SI{1.4e14}{\GeV}$, models containing representations associated with $d = 6$ operators become disfavoured as they produce too much dark matter.
This is due to the presence of a \newQ with $d = 6$ representation, causing deviations from a pure EMD cosmology (cf.\ the bump in $w$ around $T = \num{e-7}\,\TBBN$ in \cref{fig:cosmo_evo_ex}) and increasing $\Omega_a h^2$.
This effectively decreases the number of models and range of $E/N$ values around $\fax \sim \SI{1.4e14}{\GeV}$, causing the dip in the maximal $|\cagg|$ value.
For larger \fax, $|\cagg|$ increases again as expected due to the growing number of possible models.}

Note that the axion band depends on \fax, which was neglected in in ref.~\cite[][Fig.~4]{2107.12378}.
This effect is most prominent when considering the maximum possible value since going to higher \fax (lower \ma) effectively makes the LP criterion~\cref{criterion:lp} more permissive.
However, for the central values of the KSVZ axion band, the effect is only mild.

One potentially desirable property of axion models, which would avoid the introduction of additional symmetries or finetuning, is condition~\cref{des:dw}, which avoids the DW problem.
We show the band resulting from the $\NDW = 1$ subset of models in the right panel of \cref{fig:ksvz_band_cag}.
The central band for this subset lies above the band of all models, making it particularly interesting as targets for experimental searches (see \cref{sec:experiments}).

Relevant for the required experimental sensitivity is also the smallest possible value of $\cagg$.
While an accidental cancellation, $\cagg = 0$, has zero probability measure, \cagg can still be highly suppressed, leading to effectively ``photophobic'' models.
In \cref{fig:ksvz_band_cag}, we show the minimal (maximal) value of $|E/N - 1.92|$ as a solid, black line to benchmark smallest (largest) possible value of $|\cagg|$.
In particular, note that there exists a lower analytical limit on $|\cagg|$ for models with $\NDW = 1$ due to quantisation conditions~\cite{Choi:2023pdp,Reece:2023iqn}, explaining the constant horizontal line in the right panel of \cref{fig:ksvz_band_cag}.
Moreover, the quantisation conditions for the charges also allow us to perform a nontrivial consistency check on our catalogues: we verified that the anomaly coefficients in our catalogues satisfy $2N \in \mathbb{Z}$, $3E \in \mathbb{Z}$, and $4N + 3E \in 3\mathbb{Z}$~\cite{Choi:2023pdp,Reece:2023iqn} for all models.

\begin{figure*}
    \centering
    \includegraphics[width=6in]{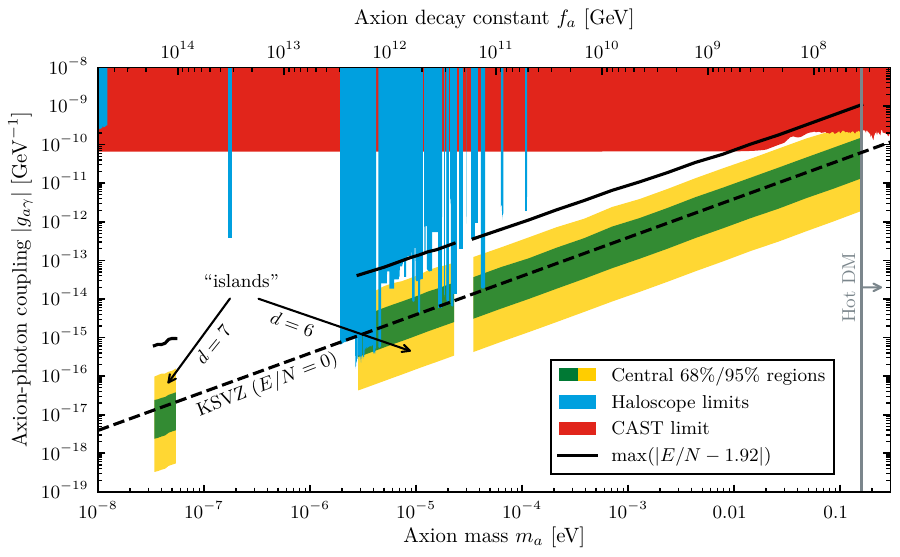}
    \caption{Self-consistent 68\% (green) and 95\% (yellow) central regions of the hadronic axion band in the post-inflationary PQ symmetry breaking scenario for $|\gagg|$. The dashed and solid lines mark the typical location of models with $E/N = 0$ and the largest possible value of $|E/N - 1.92|$, respectively. For context, we include various haloscope~(blue)~\cite{DePanfilis:1987dk,Wuensch:1989sa,Hagmann:1990tj,Hagmann:1996qd,ADMX:2009iij,Brubaker:2016ktl,McAllister:2017lkb,ADMX:2018gho,ADMX:2018ogs,HAYSTAC:2018rwy,Alesini:2019ajt,ADMX:2019uok,Lee:2020cfj,Jeong:2020cwz,HAYSTAC:2020kwv,CAST:2020rlf,CAPP:2020utb,Alesini:2020vny,ADMX:2021nhd,Grenet:2021vbb,Bartram:2021ysp,Yoon:2022gzp,Alesini:2022lnp,Lee:2022mnc,Kim:2022hmg,Yi:2022fmn,Quiskamp:2022pks,Adair:2022rtw,TASEH:2022vvu,Quiskamp:2023ehr,Yang:2023yry,HAYSTAC:2023cam,QUAX:2023gop,Kim:2023vpo,2402.19063,CAPP:2024dtx,HAYSTAC:2024jch,Ahyoune:2024klt,Quiskamp:2024oet,Bartram:2024ovw} and CAST helioscope~(red) limits~\cite{hep-ex/0702006,1705.02290,2406.16840}, provided by ref.~\cite{2020_Zenodo_OHare}, and bounds from hot DM~\cite{2310.08169}.}
\label{fig:ksvz_band_gagamma}
\end{figure*}

To better assess the interplay between theory and experimental searches, we show the more familiar KSVZ model band for $|\gagg|$ in \cref{fig:ksvz_band_gagamma}, following the same strategy for sampling the band as in \cref{fig:ksvz_band_cag}.
In particular, we include bounds from various haloscopes~\cite{DePanfilis:1987dk,Wuensch:1989sa,Hagmann:1990tj,Hagmann:1996qd,ADMX:2009iij,Brubaker:2016ktl,McAllister:2017lkb,ADMX:2018gho,ADMX:2018ogs,HAYSTAC:2018rwy,Alesini:2019ajt,ADMX:2019uok,Lee:2020cfj,Jeong:2020cwz,HAYSTAC:2020kwv,CAST:2020rlf,CAPP:2020utb,Alesini:2020vny,ADMX:2021nhd,Grenet:2021vbb,Bartram:2021ysp,Yoon:2022gzp,Alesini:2022lnp,Lee:2022mnc,Kim:2022hmg,Yi:2022fmn,Quiskamp:2022pks,Adair:2022rtw,TASEH:2022vvu,Quiskamp:2023ehr,Yang:2023yry,HAYSTAC:2023cam,QUAX:2023gop,Kim:2023vpo,2402.19063,CAPP:2024dtx,HAYSTAC:2024jch,Ahyoune:2024klt,Quiskamp:2024oet,Bartram:2024ovw}
and the CAST helioscope~\cite{hep-ex/0702006,1705.02290,2406.16840}.
Since we consider the cosmological implications of these new models, we also include additional cosmological bounds from axion hot DM, corresponding to $\ma < \SI{0.16}{\eV}$~\cite{2310.08169} (or $\ma < \SI{0.18}{\eV}$~\cite{2212.11926}) at the upper 95\% credible interval.
These cosmological bounds are driven by CMB constraints on the overproduction of dark radiation, which are approximately obtained from the bound on $\Delta \mtx{N}{eff}$ (see refs.~\cite{2101.10330,2108.04259,2211.03799,2211.05073} for recent computational advances in axion thermal production).
We do not explore these thermal axion bounds in detail, but a discussion in the context of axions and nonstandard cosmologies can be found in refs.~\cite{2308.01352,2310.16087,2411.17320}.

\subsection{Impact on axion searches}\label{sec:experiments}

Similar to ref.~\cite{2107.12378}, we find that a large part of the model space is in reach of present and future experiments.
In particular, when considering models with the desirable property~\cref{des:dw}, $\NDW = 1$, we see that these lead to generally higher values of $|\cagg|$.
Across all \fax values considered, we find that between 3\% and 12\% of models have $\NDW = 1$.

As already mentioned in \cref{sec:multipleQs}, apart from typical values of \cagg, we may also be worried about an accidental cancellation or suppression of $|\cagg|$.
When defining such ``photophobic models'' as models with $|E/N - 1.92| < 0.04$, as proposed in ref.~\cite{2107.12378}, we find that only between 1\% and 2\% of models fall into this category.

Beyond the probabilistic interpretation of the KSVZ model band, the updated catalogues contain more benchmark models -- and $E/N$ values -- than the previous catalogue~\cite{Zenodo_KSVZCatalogue}, which provide experimental benchmark targets and could connect a possible axion detection to the underlying UV physics.

Overall, our results are most relevant for existing haloscope searches around the $d = 6$ window with masses $\ma \sim \SI{10}{\ueV}$. 
This is because, in the post-inflationary scenario, these searches would typically not be sensitive to standard QCD axion models.
However, when taking into account the self-consistent EMD for KSVZ axion models, such windows will open up.
As we discuss in \cref{sec:caveats}, the exact size of the window depends on the contribution of topological defects, which can shrink the size of the window.
The haloscope signal also depends on the local axion abundance, which one should use to rescale the haloscope limits in \cref{fig:ksvz_band_gagamma}.
Despite these caveats, a haloscope discovery in one of the islands could provide evidence for a nonstandard cosmology in the post-inflationary picture, for which KSVZ models with an EMD cosmology are minimal candidates, or the pre-inflationary PQ breaking scenario.
In the former case, one may infer (limits on) the value of $\Ldec \leq \MP$, which could imply the existence of new physics at that scale, both beyond the SM and PQ scale.
As such, it still makes sense to search \emph{between} the islands.
For instance, finding a signal at $\ma \sim \SI{1}{\ueV}$ is consistent with an enlarged $d = 7$ island from setting $\Ldec \sim \SI{e17}{\GeV}$.
We discuss the effect of different choices of \Ldec in the following section.

\subsection{Caveats and limitations}\label{sec:caveats}

Although the possibility of a period of EMD in KSVZ axion models, as pointed out in ref.~\cite{2310.16087}, is quite generic, the extension of the catalogue of hadronic axion models depends on the assumptions of the KSVZ model building.
In this sense, the extended catalogue and results presented in \cref{sec:results}, and summarised in \cref{sec:conclusions}, are not as universal as for models related to standard cosmology.

\paragraph{\updated{Threshold effects.}} Similar to previous works, our results depend on the choice of \LPthresh, which can result in different catalogues and make their computation intractable.
However, while the number of viable models can change significantly, we generally find that at least the central regions of the KSVZ model band do not change much for different choices of \LPthresh.

The situation is somewhat different for the new energy scale that appears in this work, the EFT cut-off \Ldec for the decay operators in \cref{eq:decay_ops}.
Changing the scale can drastically alter the number of representations and, more importantly, the location and size of the model islands, which we approximately derived in \cref{sec:oneQ}.
Lowering \Ldec from its current value of $\Ldec = \MP$ could enlarge and shift the islands, significantly changing the qualitative nature of our results.
\updated{For instance, the $d = 6$ island will join the region for standard cosmology when lowering $\Ldec \approx 0.7\,\MP$, i.e., a value only slightly smaller than the choice adopted in this work.
Similarly, one would expect the $d = 7$ to join the $d = 6$ island when $\Ldec \sim 0.01\,\MP$.
However, for EMD cosmologies, the choice of \Ldec will also affect $\Omega_a h^2$ in a nontrivial way that depends on the other scales involved (cf.\ Ref.~\cite{0912.0015}).
Numerical investigation reveals that this only happens when the $d = 7$ merges into the standard region, which happens at $\num{7e-4}\,\MP$.}

\updated{The interplay between \Ldec and $\Omega_a h^2$ has another consequence: representations related to $d \geq 8$ can become viable if \Ldec is lowered enough.
We find that this happens for $d = 8$ operators when $\Ldec \lesssim 0.1\,\MP$, and for $d = 9$ operators when $\Ldec \lesssim 0.01\,\MP$, while $\fax \lesssim \SI{5.6e14}{\GeV}$ in both cases.}

\paragraph{\updated{Topological defects.}} Another simplification is that we neglect the contribution of cosmic strings and DWs, as mentioned in criterion~\cref{criterion:dm}.
Accurately estimating the contributions from topological defect decays to $\Omega_a h^2$ is challenging in standard cosmology, let alone in an EMD cosmology.
Still, $\Omega_a h^2$ should increase for any value of \fax when including the additional contribution from topological defects, meaning that our DM constraints on \fax would become more stringent.

\updated{For instance, we found in \cref{sec:realignment} that $\fax \lesssim \SI{2e11}{\GeV}$ in standard cosmology to satisfy the DM relic density constraint.
The analysis of cosmic string simulations, which are plagued by systematic effects and technical difficulties, suggest that $\fax \lesssim \SI{e10}{\GeV}$~\cite{2007.04990}, $\fax \lesssim \SI{e10}{\GeV}$ or $\fax \lesssim \SI{6e10}{\GeV}$~\cite{2401.17253}, or $\fax \lesssim \SI{e11}{\GeV}$~\cite{2412.08699}.
The effect of the string-DW network is yet more difficult to estimate, but there might be evidence to suggest that $\fax \lesssim \SI{2e10}{\GeV}$~\cite{2412.08699}.
In any case, the literature suggests that the standard region could be smaller by a factor between $\mathcal{O}(2)$ and $\mathcal{O}(200)$ as the lower limit of \ma is shifted upwards in \cref{fig:ksvz_band_gagamma}.
While the scaling is different for nonstandard cosmologies, which could also affect the evolution of cosmic strings, it is clear that both the $d = 7$ and the $d = 6$ regions may disappear altogether due to DM constraints in the most extreme case.
While this would be a drastic change, it could play a crucial role in better understanding the relevance of the topological defect contribution in case of an axion discovery (see \cref{sec:conclusions}).}

\updated{In this context, we also mention that many of the newly added models have $\NDW > 1$.
While the scaling of the energy density of the strings with \NDW is known, the overall power scaling of $\Omega_a h^2$ depends on the shape of the axion emission spectrum~\cite{2007.04990}.
Models with $\NDW>1$ also require additional mechanisms to make them viable at all.
Overall, however, we can assume that models with higher \NDW will lead to larger $\Omega_a h^2$ and are thus more restricted.
Since we find models with $\NDW = 1$ for all viable \fax, this will not change the extent of the standard region or the islands.
However, the model catalogues and axion band for $|\cagg|$ might change when topological defects are considered, since $\Omega_a h^2$ depends on \NDW.}

\updated{In any case, neglecting topological defects renders our publicly available model catalogues~\cite{Zenodo_UpdatedKSVZCatalogue} most conservative: they contain the maximal set of possible models in the post-inflationary PQ breaking scenario.
A reader who wishes to rescale the $\Omega_a h^2$ estimate with results from their favourite topological defect simulation, possibly involving a dependence on \NDW, may trivially do so and update the hadronic axion band accordingly.}

In the complementary, pre-inflationary PQ symmetry breaking scenario, we need not worry about topological defects, as these are likely inflated away.
Unfortunately, $\Omega_a h^2$ becomes a probabilistic quantity and cannot be computed uniquely, and the dilution could also apply to the \newQs, removing the requirement of \newQ decays and opening up the space of possible representations.
Moreover, we have to consider isocurvature constraints, which would introduce a model dependence on the inflationary model (see refs.~\cite{hep-th/0409059,0903.4377,0912.0015} for related discussions).
Note that the limit on the maximum possible reheating temperature, discussed in \cref{sec:emd}, may rule out the post-inflationary scenario for the $d = 7$ island.
 
\paragraph{\updated{Confinement.}} A further neglected effect is the possible hadronisation of \newQ states -- i.e.\ \newQ mesons and hadrons -- below the QCD crossover temperature $\mtx{T}{QCD} \approx \SI{160}{\MeV}$.
This could have a sizeable effect on the annihilation cross section in \cref{eq:sigv} for geometrical reasons 
(see e.g.\ ref.~\cite{hep-ph/0611322}). 
However, as argued in ref.~\cite{2310.16087}, in the period of 
\newQ~domination, the most likely hadrons which are formed are 
of $\bar{\newQ} \newQ$ type, thus continuing the period of EMD.

\paragraph{\updated{Small deviations of physical scales.}} Regarding the choice of \TBBN which, together with the choice of requirement on the value of $w(\TBBN)$ or performing a more detailed analysis of BBN, can have an impact on whether a model is allowed or not.
However, compared to the other choices of scales, these approximations and choices should be less important, in particular since the even simpler arguments comparing $\Gamma \sim H$ to locate the islands in \cref{sec:oneQ} seem to work rather well. 

Another such point can be made with regards to the choice of $\mQ = \fax$, implying fixing the related Yukawa coupling $y_\newQ$, and that all \newQs have an identical value of their mass \mQ.
The effect of this choice is somewhat degenerate with the choice of \Ldec, but potentially less impactful as long as $y_\newQ \sim \mathcal{O}(1)$.
However, larger variation may also lead to the disappearance of the $d = 7$ island.

Another small effect is the mild model dependence of the annihilation cross section in \cref{eq:sigv}, which seems to only lead to an $\mathcal{O}(1)$ variation of \sigv.

Similarly, as already discussed in ref.~\cite{2310.16087}, one could consider a branching ratio of the decays to non-SM states.
This would reduce the effect of decays in \cref{eq:boltzmann:te} and alter the temperature scaling of the alternative cosmology.
Still, the overall EMD cosmology would stay intact, making such an investigation only relevant in the context of hot DM.

To further simplify the cosmological computations, we ignored threshold effects around $T \sim \mQ$ regarding \gRho and the running of \alphaS.
However, as argued before, we are not interested in an accurate computation of $H(T)$ at such high scales since it is inconsequential for the computation of $\Omega_a h^2$.

\paragraph{\updated{Higher-order effects.}} Also recall that, for each \newQ, we only include terms related to the lowest-dimensional decay operator compatible with the \newQ's representation.
Given that the next-highest operator are suppressed by a factor of order $(\mQ/\Ldec)^2$, we deem this an excellent approximation, which helped us to further reduce the model dependence of the results since the compatibility of higher-dimensional operators would have to be checked on a case-by-case basis.

\section{Summary and conclusions}\label{sec:conclusions}

We extended the catalogue of \pref hadronic (KSVZ) axion models to include cosmologically viable representations associated with higher-dimensional ($d \geq 6$) decay operators.
As pointed out in ref.~\cite{2310.16087}, such higher-dimensional operators can lead to a phase of EMD, alter the cosmological history, and consequently relax the constraints on \newQ decays.
Compared to the previous work, which only considered KSVZ models with one new, heavy fermion \newQ, we now consider multiple \newQs.
Our work also goes beyond a purely phenomenological study by -- for the first time -- studying the underlying model building in terms of the \newQ representations.
Compared to the previous study of the hadronic axion model band~\cite{2107.12378}, we now consider the model band's dependence on~\fax.

Depending on \fax, we found between \num{44292} and \num{99877479} nonequivalent models and between \num{262} and \num{2035} unique $E/N$ values compared to the previously identified \num{820} different $E/N$~values at $\fax = \SI{5e11}{\GeV}$~\cite{2107.12378}.
However, it should be stressed that the numerical values in the extended catalogue are more dependent on the underlying assumptions and choices than in the previous catalogues due to the importance of the cosmological evolution (see \cref{sec:caveats}).
Changing some or all of these choices can potentially result in much larger or smaller catalogues.
Let us therefore stress the more generic results of our work:
\begin{itemize}
    \item The LP criterion, in addition to BBN and DM constraints, can be used to extend the KSVZ model catalogue to models whose representations are compatible with decay operators up to dimension $d \leq 7$.
    \item While the contents of the model catalogue \updated{may} change significantly depending on \updated{the adopted cutoff values} in this work, we find that the central band of hadronic axion models is a fairly stable prediction\updated{, providing robust targets for axion searches.}
    \item We expect viable model islands around $\fax \sim \SI{e12}{\GeV}$ and $\fax \sim \SI{e14}{\GeV}$ in the post-inflationary PQ symmetry breaking scenario, whose size \updated{and location are subject to the adopted} assumptions and the size of the contribution from topological defects.
\end{itemize}
While one could have already anticipated the existence of the model islands from ref.~\cite{2310.16087}, we determined their boundaries in post-inflationary PQ symmetry breaking scenario for the entire catalogue of hadronic models.
More generally, the extended KSVZ model database and band helps us to provide theory-inspired targets for experimental searches and updates the definition of the QCD~axion model band.

The new models can lie beyond the standard post-inflationary QCD~axion mass window, which -- on the one hand -- complicates the landscape of model but -- on the other hand -- presents an opportunity for conducting phenomenologically-driven axion searches around these parameter regions.
In particular, detecting an axion in \updated{or between} the model islands could teach us more about the topological defect contribution to the axion abundance, the underlying UV model, or \updated{a possible} nonstandard cosmological scenario (see ref.~\cite{1906.04183} for reconstructing nonstandard cosmologies with weakly-interacting massive particles).

To facilitate \updated{such use cases}, we make our updated catalogues, summary histograms, and related analysis scripts available on Github~\cite{Github_UpdatedKSVZCatalogue} (version tag v1.0) and Zenodo~\cite{Zenodo_UpdatedKSVZCatalogue}.
\updated{By ignoring a possible topological defect contribution and making conservative parameter choices where possible, we took care not to exclude potentially viable models.
The supplemental material allows others to recalibrate the axion model bands using other assumptions and parameters in the future.}

\begin{acknowledgments}
We thank Andrew Cheek, Jordy de~Vries, Ken'ichi Saikawa, and Wen Yin for helpful comments.
The work of LDL is supported by the European Union -- Next Generation EU and by the Italian Ministry of University and Research (MUR) via the PRIN 2022 project n.~2022K4B58X -- AxionOrigins.
SH has received funding from the European Union's Horizon Europe research and innovation programme under the Marie Sk{\l}odowska-Curie grant agreement No~101065579. VP thanks the University of Padova and INFN Padova for their hospitality.
\end{acknowledgments}

\appendix

\section{Perturbative unitarity bounds on the photophilic axion}
\label{sec:unitarity}

Consider the axion EFT 
\begin{equation}
\mathcal{L} = C_{a\gamma} \frac{\alphaEM}{8\pi} \frac{a}{f_a} F \tilde F 
+ C_{a\psi} \frac{\partial_\mu a}{2 f_a} \overline{\psi} \gamma^\mu \gamma_5 \psi + \dots \, ,
\end{equation}
with $\psi$ denoting a generic SM fermion. Here $C_{a\gamma} = E/N - 1.92 \simeq E/N$ is assumed to be $\gg 1$, while $C_{a\psi}$ is of $\mathcal{O}(1)$.
We can estimate the validity range of the axion EFT by requiring perturbative unitarity of axion-mediated scattering amplitudes.
To this end, consider $2\to 2$ scattering of fermions and photons, 
which yield
\begin{equation}
\mathcal{M}_{\psi \psi \to \psi \psi} \sim 
C_{a\psi}^2\frac{s}{f_a^2} \quad \text{and} \quad
\mathcal{M}_{\gamma\gamma \to \gamma\gamma} \sim 
\left(
\frac{\alphaEM}{8\pi}
\right)^2
C_{a\gamma}^2\frac{s}{f_a^2} \, ,  
\end{equation}
where $s$ denotes the usual Mandelstam variable.
The bounds from perturbative unitarity are 
respectively of the order of
\begin{equation}
\sqrt{s} \lesssim \sqrt{4 \pi} \frac{f_a}{C_{a\psi}} \simeq \sqrt{4 \pi} f_a \quad \text{and} \quad
\sqrt{s} \lesssim \sqrt{4 \pi} \frac{8\pi}{\alphaEM} \frac{f_a}{C_{a\gamma}} \simeq \sqrt{4 \pi} \frac{8\pi}{\alphaEM} \frac{f_a}{E/N} \, .
\end{equation}
Hence, while fermion-fermion scattering implies that the axion EFT should 
be UV-completed at energies of order $\sqrt{4\pi} f_a$, the bound from 
photon-photon scattering suggests instead that the axion EFT breaks down at
\begin{equation}
E_{\star} \sim \sqrt{4 \pi} \frac{8\pi}{\alphaEM} \frac{f_a}{E/N} \approx
3.4 \, \sqrt{4 \pi} \, \fax \left( \frac{10^3}{E/N} \right) \, .
\end{equation}
Note that, for $E/N \gtrsim 10^3$, the axion EFT breaks down 
earlier than the {na\"ive} expectation of a breakdown at $E_{\star} \sim \sqrt{4\pi} \fax$. 
In the extreme case of $E/N \gg 10^3$, which could be realised in clockwork-like models \cite{Farina:2016tgd,Darme:2020gyx} or by relaxing LP constraints in KSVZ models, the cut-off scale may be much lower than \fax, potentially endangering the calculability in the axion EFT below \fax.

\section{Identifying new decay operators with \deco}\label{sec:deco}

The code \deco~\cite{2212.04395} is a \form~\cite{Vermaseren:2000nd,Ruijl:2017dtg} implementation of the Hilbert series~\cite{Lehman:2015via,Lehman:2015coa,Henning:2015daa}, designed to create subleading operator bases for effective theories with freely chosen field content. It supports symmetry groups typically encountered in flavour and beyond-the-SM physics. 
The main mathematical expressions needed to enumerate operators with the Hilbert series are the \emph{characters} of the group representations involved (see, e.g., ref.~\cite{Banerjee:2020bym} for an overview). Characters are the traces of the group elements, and for Lie groups of rank $r$, they are given by functions of $r$ complex variables. 
Characters of group representations exhibit two properties that are essential for the enumeration of operators.
First, the character of a product of two representations equals the sum of characters of the irreducible representations contained in the product representation.
Second, the characters of irreducible representations are orthogonal, satisfying: 
\begin{equation}\label{eq:characterOrthogonality}
\int d\muH\ \chi_R\ \chi_{R'}^* = \delta_{RR'}\,,
\end{equation}
where $\chi_R$ and $\chi_{R'}$ are the characters of two irreducible representations $R$ and $R'$, and \muH is the Haar measure associated with the group in question.
The star in $\chi_{R'}^*$ indicates complex conjugation.
The Haar measures for the SM gauge groups are
\begin{align}
\int d\mu_{\gr{U}(1)}=\frac{1}{2\pi\ii} \oint_{|z|=1} \! \frac{\dd z}{z} \,,
\qquad
\int d\mu_{\gr{SU}(2)}=\frac{1}{2\pi\ii} \oint_{|z|=1}\! \frac{\dd z}{z} \; (1-z^2)\,, \nonumber \\
\int d\mu_{\gr{SU}(3)}=  \frac{1}{(2\pi\ii)^2} \oint_{|z_1|=1} \! \frac{\dd z_1}{z_1} \; \oint_{|z_2|=1} \!\frac{\dd z_2}{z_2} \; \left(1-z_1 z_2\right) \left(1-\frac{z_1^2}{z_2}\right) \left(1- \frac{z_2^2}{z_1}\right)\,.
\end{align}
These properties allow us to project out the number of singlets contained in a product representations by substituting one character with trivial character ``1'' in \cref{eq:characterOrthogonality}.
The full picture is of course more complicated due to additional constraints on the operator basis: bosonic fields commute and fermions anti-commute. Furthermore, we discard operators related through the equations of motions and total derivatives.
More details on how these constraints are implemented in the Hilbert series can be found in ref.~\cite{Henning:2017fpj}. 

One of the key features of \deco is that the code is highly modular and can easily be extended to additional groups and/or representations.
For this work, this modularity was utilised to extend the existing $\gr{SU}(2)$ and $\gr{SU}(3)$ representations.
The relevant representations for this work, tabulated in \cref{tab:characters} together with the relevant characters, are now included in version~1.1 of \deco.
As discussed in the main text, the charges of the \newQs are \textit{a~priori} undetermined, necessitating a scan over all possible charges.
To accommodate this, \deco was further adapted such that the main procedure for the enumeration of operators can be executed multiple times in a loop within the same \form program.
To allow others to reproduce our results, we provide the corresponding \form input files for \deco on Zenodo~\cite{Zenodo_UpdatedKSVZCatalogue}.

\begin{table}
    \centering
    \caption{Characters of representations relevant for this work. The characters of conjugate representations in $\gr{SU}(3)$ can be obtained via the replacement $z_1 \leftrightarrow z_2$.}\label{tab:characters}
    \small
    \renewcommand{\arraystretch}{1.75}
    \begin{tabular}{lcc}
    \toprule
    Group & Rep. & Character\\
    \midrule
    
    $\gr{U}(1)$ & $Q$ & $z^Q$ \\
    
    \midrule
    
    $\gr{SU}(2)$ & 1 & 1  \\
     & \ccol  $n$ & \ccol $\sum_{m \in M} z^{2m} \quad \text{with} \quad M = \left\{-\frac{(n-1)}{2},\,-\frac{(n-1)}{2}+1,\,\dots,\frac{n-1}{2}\right\}$ \\
    
    \midrule

    $\gr{SU}(3)$ & 1 & 1 \\
    & \ccol 3 & \ccol $z_1 + \frac{z_2}{z_1} + \frac{1}{z_2}$ \\
    & 6 & $z_1^2+\frac{z_1}{z_2}+z_2+\frac{1}{z_2^2}+\frac{1}{z_1}+\frac{z_2^2}{z_1^2}$  \\
    & \ccol 8 & \ccol $\frac{z_1^2}{z_2}+z_2 z_1+\frac{z_1}{z_2^2}+\frac{z_2^2}{z_1}+\frac{1}{z_2 z_1}+\frac{z_2}{z_1^2}+2$ \\
    & 10 & $z_1^3+\frac{z_1^2}{z_2}+z_2 z_1+\frac{z_1}{z_2^2}+\frac{1}{z_2^3}+\frac{z_2^2}{z_1}+\frac{1}{z_2 z_1}+\frac{z_2}{z_1^2}+\frac{z_2^3}{z_1^3}+1$ \\
    & \ccol 15 & \ccol $\frac{z_1^3}{z_2}+z_2 z_1^2+\frac{z_1^2}{z_2^2}+\frac{z_1}{z_2^3}+2 z_1+z_2^2+\frac{2}{z_2}+\frac{2 z_2}{z_1}+\frac{1}{z_2^2 z_1}+\frac{z_2^3}{z_1^2}+\frac{1}{z_1^2}+\frac{z_2^2}{z_1^3}$ \\
    & $15'$ & $z_1^4+\frac{z_1^3}{z_2}+z_2 z_1^2+\frac{z_1^2}{z_2^2}+\frac{z_1}{z_2^3}+z_1+z_2^2+\frac{1}{z_2}+\frac{1}{z_2^4}+\frac{z_2}{z_1}+\frac{1}{z_2^2
   z_1}+\frac{z_2^3}{z_1^2}+\frac{1}{z_1^2}+\frac{z_2^2}{z_1^3}+\frac{z_2^4}{z_1^4}$ \\
   & \ccol 21 & \ccol \makecell{$\frac{z_1^5}{z_2^5}+\frac{z_1^4}{z_2^3}+\frac{z_1^3}{z_2}+\frac{z_1^3}{z_2^4}+z_2 z_1^2+\frac{z_1^2}{z_2^2}+z_2^3 z_1+\frac{z_1}{z_2^3}+z_1+z_2^5+z_2^2+\frac{1}{z_2}+\frac{z_2^4}{z_1}+\frac{z_2}{z_1}+$\\ $+\frac{1}{z_2^2 z_1}+\frac{z_2^3}{z_1^2}+\frac{1}{z_1^2}+\frac{z_2^2}{z_1^3}+\frac{1}{z_2 z_1^3}+\frac{z_2}{z_1^4}+\frac{1}{z_1^5}$} \\
   & 24 & \makecell{$\frac{z_1^4}{z_2^3}+\frac{z_1^3}{z_2}+\frac{z_1^3}{z_2^4}+z_2 z_1^2+\frac{2 z_1^2}{z_2^2}+z_2^3 z_1+\frac{z_1}{z_2^3}+2 z_1+2 z_2^2+\frac{2}{z_2}+\frac{z_2^4}{z_1}+\frac{2 z_2}{z_1}+\frac{1}{z_2^2 z_1}+$\\ $+\frac{z_2^3}{z_1^2}+\frac{2}{z_1^2}+\frac{z_2^2}{z_1^3}+\frac{1}{z_2 z_1^3}+\frac{z_2}{z_1^4}$} \\
   & \ccol 27 & \ccol \makecell{$\frac{z_1^4}{z_2^2}+\frac{z_1^3}{z_2^3}+z_1^3+z_2^2 z_1^2+\frac{2 z_1^2}{z_2}+\frac{z_1^2}{z_2^4}+2 z_2 z_1+\frac{2 z_1}{z_2^2}+z_2^3+\frac{1}{z_2^3}+\frac{2 z_2^2}{z_1}+\frac{2}{z_2 z_1}+$\\ $+\frac{z_2^4}{z_1^2}+\frac{2 z_2}{z_1^2}+\frac{1}{z_2^2 z_1^2}+\frac{z_2^3}{z_1^3}+\frac{1}{z_1^3}+\frac{z_2^2}{z_1^4}+3$} \\
   & 28 & \makecell{$z_1^6+\frac{z_1^5}{z_2}+z_2 z_1^4+\frac{z_1^4}{z_2^2}+\frac{z_1^3}{z_2^3}+z_1^3+z_2^2 z_1^2+\frac{z_1^2}{z_2}+\frac{z_1^2}{z_2^4}+z_2 z_1+\frac{z_1}{z_2^2}+\frac{z_1}{z_2^5}+\frac{1}{z_2^3}+1+z_2^3+$\\ $+\frac{1}{z_2^6}+\frac{z_2^2}{z_1}+\frac{1}{z_2 z_1}+\frac{1}{z_2^4 z_1}+\frac{z_2^4}{z_1^2}+\frac{z_2}{z_1^2}+\frac{1}{z_2^2 z_1^2}+\frac{z_2^3}{z_1^3}+\frac{1}{z_1^3}+\frac{z_2^5}{z_1^4}+\frac{z_2^2}{z_1^4}+\frac{z_2^4}{z_1^5}+\frac{z_2^6}{z_1^6}$} \\
   & \ccol 35 & \ccol \makecell{$\frac{z_1^5}{z_2}+z_2 z_1^4+\frac{z_1^4}{z_2^2}+\frac{z_1^3}{z_2^3}+2 z_1^3+z_2^2 z_1^2+\frac{2 z_1^2}{z_2}+\frac{z_1^2}{z_2^4}+2 z_2 z_1+\frac{2 z_1}{z_2^2}+\frac{z_1}{z_2^5}+$\\ $+z_2^3+\frac{2}{z_2^3}+\frac{2 z_2^2}{z_1}+\frac{2}{z_2 z_1}+\frac{1}{z_2^4 z_1}+\frac{z_2^4}{z_1^2}+\frac{2 z_2}{z_1^2}+\frac{1}{z_2^2 z_1^2}+\frac{2 z_2^3}{z_1^3}+\frac{1}{z_1^3}+\frac{z_2^5}{z_1^4}+\frac{z_2^2}{z_1^4}+\frac{z_2^4}{z_1^5}+2$} \\
   & 42 & \makecell{$\frac{z_1^5}{z_2^2}+\frac{z_1^4}{z_2^3}+z_1^4+z_2^2 z_1^3+\frac{2 z_1^3}{z_2}+\frac{z_1^3}{z_2^4}+2 z_2 z_1^2+\frac{2 z_1^2}{z_2^2}+\frac{z_1^2}{z_2^5}+z_2^3 z_1+\frac{2 z_1}{z_2^3}+3 z_1+\frac{3}{z_2}+$\\ $+2 z_2^2+\frac{1}{z_2^4}+\frac{z_2^4}{z_1}+\frac{3 z_2}{z_1}+\frac{2}{z_2^2 z_1}+\frac{2 z_2^3}{z_1^2}+\frac{1}{z_2^3 z_1^2}+\frac{2}{z_1^2}+\frac{z_2^5}{z_1^3}+\frac{2 z_2^2}{z_1^3}+\frac{1}{z_2 z_1^3}+\frac{z_2^4}{z_1^4}+\frac{z_2}{z_1^4}+\frac{z_2^3}{z_1^5}$} \\
   \bottomrule
   \end{tabular}
\end{table}

\section{Analytical approximation for the allowed representation sizes}
\label{sec:analyticalapprox}

The LP criterion \cref{criterion:lp} is the most restrictive one in constraining the size of the \newQ representations (and thus the number of models) that fall in the \pref window.
Although we numerically compute the two-loop beta functions to restrict the model space, it is helpful to look at the running of the couplings at one loop to understand the numerical results using an analytical approximation.

At the one-loop level, the running of the SM gauge couplings $\alpha_i = g_i^2/(4\pi)$ are given by
\begin{align}
    \frac{\dd}{\dd\mathrm{t}}\alpha_i^{-1} &= -a_i  \, \label{eq:rge} \, ,
\end{align}
where $i \in \{1,\,2,\,3\}$, $\mathrm{t} = \frac{1}{2\pi}\ln(\mu/m_Z)$ for energy scale $\mu$ and $Z$~boson mass $m_Z$, and
\begin{align}
    a_i &= -\frac{11}{3} \, C_2(G_i) + \frac{4}{3}\sum_F \kappa \, \dynk(F_i) + \frac{1}{3} \sum_S \eta \, \dynk(S_i)
    \label{eq:betafns}
\end{align}
being the one-loop beta functions.
$C_2$~and $\dynk$ denote the quadratic Casimir and Dynkin indices of the corresponding gauge group, respectively, for the fermionic $(F)$ and scalar $(S)$ fields in the theory, while $G_i$ denotes the adjoint representation of the gauge group. $\kappa=\frac{1}{2},1$ for Weyl and Dirac fermions, and $\eta=1$ for complex scalars. For the full two-loop expression we use in this work, see, e.g., refs.~\cite{2107.12378,Machacek:1983tz,1504.00359}.

\begin{figure}
    \centering
    \includegraphics[width=0.65\linewidth]{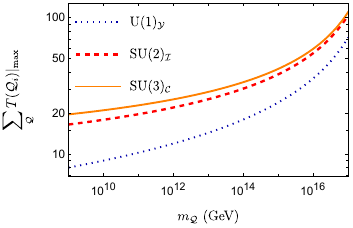}
    \caption{The maximum value of the sum of Dynkin indices of the additional \newQs allowed by the LP criterion from the one-loop beta functions. We show the results for the three SM gauge groups as different lines.}
    \label{fig:analytical_limit}
\end{figure}
It is clear from \cref{eq:rge} that at one loop, the couplings do not mix, and \LLP is easily determined as a function of the mass \mQ from the beta functions.
Demanding that $\LLP \geq \MP$, we can then set a limit on $(a_\newQ)_i \equiv a_i - (a_\text{SM})_i$, the difference in the one-loop beta function caused by the addition of \newQ to the SM, as a function of $\mQ$, which in turn gives us an upper bound on the sum of Dynkin indices related to the additional new quarks, as shown in \cref{fig:analytical_limit}.

The results in \cref{fig:analytical_limit} provide an estimate of the largest possible size of \newQ representations.
This estimate is approximate since the couplings only run independently of each other at one loop, while the two-loop beta functions introduce mixings.
At one loop, each group is, in effect, agnostic to the \newQ representation in the other groups, i.e., the effect would be the same as \newQs transforming trivially under the other groups.
However, it turns out that the {na\"ive} results in \cref{fig:analytical_limit} are close to the full numeric treatment employed for the creation of catalogues in this work, but a bit weaker -- not only due to the fact that we use two-loop beta functions in the main text, but also because the \newQs must necessarily be colour-charged, which accelerates the running of all couplings, and they consequently already hit a LP at smaller energies compared to the one-loop results.

\section{Complete list of allowed representations}

In \cref{tab:all_reps} we give a list of all the \newQ representations surviving the LP criterion, for $\NQ = 1$, setting $\mQ = \SI{e17}{\GeV}$, and allowing operators up to mass dimension $d=9$.
The first 20 entries, i.e.\ for $d\leq5$, were already found and studied in refs.~\cite{1610.07593,1705.05370,2107.12378}.
Note that several of these representations will be disallowed, especially those with larger minimum $d$ as they tend to have large SM charges, for smaller values of $\mQ$ since the LP will be induced earlier.
Some of the representations can also jeopardise cosmological consistency as detailed in \cref{sec:cosmo}, and the number of models can be further reduced.

\begin{center}
\begin{longtable}{rS[table-format=2]S[table-format=-2]ccccS[table-format=1.1e2]}
  \caption{Allowed \newQ representations (by the LP criterion) for $\NQ = 1$ and $\mQ = \SI{e17}{\GeV}$, when considering decay operators up to dimension $d = 9$. The first three columns give the SM charges; the fourth and fifth columns give the corresponding anomaly ratio and the domain wall number, assuming no other source of such anomalies exist; the sixth column shows the lowest mass dimension at which the representation can appear; the seventh column gives an example operator; and the last column shows the lowest scale at which a LP appears in any of the gauge couplings.}
  \label{tab:all_reps} \\
  \toprule
  \multicolumn{3}{c}{Rep.\ $(\mathcal{C},\mathcal{I},6\mathcal{Y})$} & $E/N$ & $\NDW$ & Min.\ $d$ & Example operator & \multicolumn{1}{c}{LP [GeV]} \\
  \midrule
  \endfirsthead
  \multicolumn{7}{c}{{\tablename\ \thetable{} -- continued from previous page}} \\
  \midrule
  \multicolumn{3}{c}{Rep.\ $(\mathcal{C},\mathcal{I},6\mathcal{Y})$} & $E/N$ & $\NDW$ & Min.\ $d$ & Example operator & \multicolumn{1}{c}{LP [GeV]} \\
  \midrule
  \endhead
  \midrule
  \multicolumn{7}{r}{{Continued on next page}} \\
  \midrule
  \endfoot
  \bottomrule
  \endlastfoot
    3 & 1 & -2 & 2/3 & 1 & 3 & $\bar{\newQ}_L\,d_R$ & 2.0e+39 \\
   3 & 1 & 4 & 8/3 & 1 & 3 & $\bar{\newQ}_L\,u_R$ & 6.8e+35 \\
   3 & 2 & 1 & 5/3 & 2 & 3 & $\bar{\newQ}_R\,q_L$ & 9.7e+39 \\
   \midrule
   3 & 2 & -5 & 17/3 & 2 & 4 & $H^\dagger\,\bar{\newQ}_L\,d_R$ & 1.1e+30 \\
   3 & 2 & 7 & 29/3 & 2 & 4 & $H\,\bar{\newQ}_L\,u_R$ & 1.1e+26 \\
   3 & 3 & -2 & 14/3 & 3 & 4 & $H^\dagger\,\bar{\newQ}_R\,q_L$ & 6.2e+36 \\
   3 & 3 & 4 & 20/3 & 3 & 4 & $H\,\bar{\newQ}_R\,q_L$ & 1.7e+30 \\
   \midrule
   3 & 3 & -8 & 44/3 & 3 & 5 & $(H^\dagger)^{2}\,\bar{\newQ}_L\,d_R$ & 4.0e+22 \\
   3 & 3 & 10 & 62/3 & 3 & 5 & $(H)^{2}\,\bar{\newQ}_L\,u_R$ & 7.4e+20 \\
   3 & 4 & -5 & 35/3 & 4 & 5 & $(H^\dagger)^{2}\,\bar{\newQ}_R\,q_L$ & 2.9e+23 \\
   3 & 4 & 1 & 23/3 & 4 & 5 & $\bar{\newQ}_R\,q_L\,\sigma \cdot W$ & 3.1e+23 \\
   3 & 4 & 7 & 47/3 & 4 & 5 & $(H)^{2}\,\bar{\newQ}_R\,q_L$ & 2.7e+22 \\
   $\overline{6}$ & 1 & 4 & 16/15 & 5 & 5 & $\bar{u}_R\,\sigma \cdot G\,\newQ_L$ & 3.6e+32 \\
   $\overline{6}$ & 1 & -2 & 4/15 & 5 & 5 & $\bar{d}_R\,\sigma \cdot G\,\newQ_L$ & 9.9e+37 \\
   $\overline{6}$ & 2 & 1 & 2/3 & 10 & 5 & $\bar{q}_L\,\sigma \cdot G\,\newQ_R$ & 1.8e+39 \\
   8 & 1 & -6 & 8/3 & 6 & 5 & $\bar{e}_R\,\sigma \cdot G\,\newQ_L$ & 1.4e+26 \\
   8 & 2 & -3 & 4/3 & 12 & 5 & $\bar{\ell}_L\,\sigma \cdot G\,\newQ_R$ & 1.8e+30 \\
   15 & 1 & -2 & 1/6 & 20 & 5 & $\bar{\newQ}_L\,d_R\,\sigma \cdot G$ & 2.9e+32 \\
   15 & 1 & 4 & 2/3 & 20 & 5 & $\bar{\newQ}_L\,u_R\,\sigma \cdot G$ & 1.3e+27 \\
   15 & 2 & 1 & 5/12 & 40 & 5 & $\bar{\newQ}_R\,q_L\,\sigma \cdot G$ & 2.1e+22 \\
   \midrule
   3 & 1 & -14 & 98/3 & 1 & 6 & $\bar{\newQ}_L\,d_R\,(\bar{e}_R^c\,e_R)$ & 2.2e+22 \\
   $\overline{3}$ & 1 & 8 & 32/3 & 1 & 6 & $\bar{u}_R\,\gamma_\mu\,e_R\,\bar{d}_R\,\gamma^\mu\,\newQ_R$ & 3.0e+28 \\
   $\overline{3}$ & 1 & -10 & 50/3 & 1 & 6 & $(\bar{d}_R\,d_R^c)\,\bar{e}_R\,\newQ_L$ & 6.4e+25 \\
   3 & 1 & 16 & 128/3 & 1 & 6 & $\bar{\newQ}_L\,u_R\,(\bar{e}_R\,e_R^c)$ & 1.8e+21 \\
   3 & 2 & -11 & 65/3 & 2 & 6 & $(\bar{u}_R\,u_R^c)\,\bar{\newQ}_R\,\ell_L$ & 4.1e+21 \\
   $\overline{3}$ & 2 & -13 & 89/3 & 2 & 6 & $\bar{\ell}_L\,u_R\,\bar{e}_R\,\newQ_L$ & 2.7e+20 \\
   3 & 4 & -11 & 83/3 & 4 & 6 & $(H^\dagger)^{3}\,\bar{\newQ}_L\,d_R$ & 3.2e+19 \\
   3 & 4 & 13 & 107/3 & 4 & 6 & $(H)^{3}\,\bar{\newQ}_L\,u_R$ & 6.7e+18 \\
   3 & 5 & -8 & 68/3 & 5 & 6 & $(H^\dagger)^{3}\,\bar{\newQ}_R\,q_L$ & 7.4e+19 \\
   3 & 5 & -2 & 38/3 & 5 & 6 & $H^\dagger\,\bar{\newQ}_R\,q_L\,\sigma \cdot W$ & 8.0e+19 \\
   3 & 5 & 4 & 44/3 & 5 & 6 & $H\,\bar{\newQ}_R\,q_L\,\sigma \cdot W$ & 7.9e+19 \\
   3 & 5 & 10 & 86/3 & 5 & 6 & $(H)^{3}\,\bar{\newQ}_R\,q_L$ & 2.5e+19 \\
   6 & 1 & -10 & 20/3 & 5 & 6 & $(\bar{d}_R^c\,d_R)\,\bar{\newQ}_L\,e_R$ & 2.3e+22 \\
   6 & 1 & 8 & 64/15 & 5 & 6 & $\bar{e}_R\,\gamma_\mu\,u_R\,\bar{\newQ}_R\,\gamma^\mu\,d_R$ & 3.3e+24 \\
   $\overline{6}$ & 1 & -14 & 196/15 & 5 & 6 & $(\bar{u}_R^c\,u_R)\,\bar{e}_R\,\newQ_L$ & 9.7e+19 \\
   6 & 2 & -7 & 58/15 & 10 & 6 & $(\bar{d}_R^c\,d_R)\,\bar{\newQ}_R\,\ell_L$ & 3.4e+22 \\
   6 & 2 & 5 & 34/15 & 10 & 6 & $\bar{\ell}_L\,d_R\,\bar{\newQ}_L\,u_R$ & 8.4e+25 \\
   $\overline{6}$ & 2 & -11 & 26/3 & 10 & 6 & $(\bar{u}_R^c\,u_R)\,\bar{\ell}_L\,\newQ_R$ & 3.4e+19 \\
   6 & 3 & -4 & 8/3 & 15 & 6 & $(\bar{q}_L\,q_L^c)\,\bar{\newQ}_L\,d_R$ & 3.4e+25 \\
   6 & 3 & 2 & 28/15 & 15 & 6 & $(\bar{q}_L\,q_L^c)\,\bar{\newQ}_L\,u_R$ & 3.7e+25 \\
   6 & 3 & 8 & 88/15 & 15 & 6 & $q_L\,u_R\,\bar{\ell}_L\,\bar{\newQ}_R$ & 1.4e+20 \\
   6 & 4 & -1 & 46/15 & 20 & 6 & $(\bar{q}_L^c\,q_L)\,\ell_L\,\bar{\newQ}_R$ & 9.2e+19 \\
   $\overline{6}$ & 4 & -5 & 14/3 & 20 & 6 & $(\bar{q}_L^c\,q_L)\,\bar{\ell}_L\,\newQ_R$ & 9.0e+19 \\
   8 & 1 & 0 & 0 & 6 & 6 & $(\bar{d}_R\,d_R^c)\,\bar{u}_R\,\newQ_L$ & 5.3e+40 \\
   8 & 1 & -12 & 32/3 & 6 & 6 & $\bar{d}_R\,\gamma_\mu\,u_R\,\bar{e}_R\,\gamma^\mu\,\newQ_R$ & 1.2e+20 \\
   8 & 2 & -9 & 20/3 & 12 & 6 & $\bar{\ell}_L\,u_R\,\bar{d}_R\,\newQ_L$ & 7.0e+19 \\
   8 & 3 & 0 & 16/9 & 18 & 6 & $\bar{q}_L\,d_R\,\bar{\ell}_L\,\newQ_R$ & 1.3e+23 \\
   8 & 3 & -6 & 40/9 & 18 & 6 & $\bar{q}_L\,u_R\,\bar{\ell}_L\,\newQ_R$ & 1.1e+21 \\
   8 & 4 & -3 & 4 & 24 & 6 & $\bar{q}_L\,\gamma_\mu\,q_L\,\bar{\ell}_L\,\gamma^\mu\,\newQ_L$ & 1.5e+19 \\
   10 & 1 & 0 & 0 & 15 & 6 & $\bar{\newQ}_L\,u_R\,(\bar{d}_R^c\,d_R)$ & 4.5e+40 \\
   10 & 1 & 6 & 4/3 & 15 & 6 & $(\bar{u}_R^c\,u_R)\,\bar{\newQ}_L\,d_R$ & 6.8e+24 \\
   10 & 2 & -3 & 2/3 & 30 & 6 & $\bar{\newQ}_R\,q_L\,(\bar{d}_R^c\,d_R)$ & 6.5e+24 \\
   10 & 2 & 3 & 2/3 & 30 & 6 & $(\bar{q}_L^c\,q_L)\,\bar{\newQ}_R\,q_L$ & 6.5e+24 \\
   10 & 2 & 9 & 10/3 & 30 & 6 & $\bar{\newQ}_R\,q_L\,(\bar{u}_R^c\,u_R)$ & 2.0e+19 \\
   10 & 3 & 0 & 8/9 & 45 & 6 & $(\bar{q}_L^c\,q_L)\,\bar{\newQ}_L\,d_R$ & 2.7e+21 \\
   10 & 3 & 6 & 20/9 & 45 & 6 & $(\bar{q}_L^c\,q_L)\,\bar{\newQ}_L\,u_R$ & 1.8e+20 \\
   $\overline{15}$ & 1 & 8 & 8/3 & 20 & 6 & $(\bar{d}_R^c\,d_R)\,\bar{u}_R\,\newQ_L$ & 4.7e+20 \\
   $\overline{15}$ & 1 & -10 & 25/6 & 20 & 6 & $(\bar{u}_R^c\,u_R)\,\bar{d}_R\,\newQ_L$ & 2.9e+19 \\
   $\overline{15}$ & 2 & 5 & 17/12 & 40 & 6 & $(\bar{d}_R^c\,d_R)\,\bar{q}_L\,\newQ_R$ & 3.1e+21 \\
   $\overline{15}$ & 2 & -7 & 29/12 & 40 & 6 & $(\bar{u}_R^c\,u_R)\,\bar{q}_L\,\newQ_R$ & 3.4e+19 \\
   15 & 3 & -2 & 7/6 & 60 & 6 & $\bar{\newQ}_R\,q_L\,\bar{q}_L\,d_R$ & 8.5e+19 \\
   15 & 3 & 4 & 5/3 & 60 & 6 & $\bar{\newQ}_R\,q_L\,\bar{q}_L\,u_R$ & 8.4e+19 \\
   $\overline{15}$ & 4 & -1 & 23/12 & 80 & 6 & $(\bar{q}_L^c\,q_L)\,\bar{q}_L\,\newQ_R$ & 1.3e+18 \\
   \midrule
   3 & 2 & -17 & 149/3 & 2 & 7 & $H^\dagger\,\bar{\newQ}_L\,d_R\,(\bar{e}_R^c\,e_R)$ & 1.2e+19 \\
   3 & 2 & 19 & 185/3 & 2 & 7 & $H\,\bar{\newQ}_L\,u_R\,(\bar{e}_R\,e_R^c)$ & 4.4e+18 \\
   3 & 3 & -14 & 110/3 & 3 & 7 & $H^\dagger\,(\bar{u}_R\,u_R^c)\,\bar{\newQ}_R\,\ell_L$ & 1.2e+19 \\
   $\overline{3}$ & 3 & -16 & 140/3 & 3 & 7 & $H\,\bar{\ell}_L\,u_R\,\bar{e}_R\,\newQ_L$ & 3.9e+18 \\
   3 & 5 & -14 & 134/3 & 5 & 7 & $(H^\dagger)^{4}\,\bar{\newQ}_L\,d_R$ & 1.8e+18 \\
   3 & 6 & -11 & 113/3 & 6 & 7 & $(H^\dagger)^{4}\,\bar{\newQ}_R\,q_L$ & 3.3e+18 \\
   3 & 6 & -5 & 65/3 & 6 & 7 & $(H^\dagger)^{2}\,\bar{\newQ}_R\,q_L\,\sigma \cdot W$ & 3.6e+18 \\
   3 & 6 & 1 & 53/3 & 6 & 7 & $\bar{\newQ}_R\,q_L\,(\sigma \cdot W)^{2}$ & 3.6e+18 \\
   3 & 6 & 7 & 77/3 & 6 & 7 & $(H)^{2}\,q_L\,\sigma \cdot W\,\bar{\newQ}_R$ & 3.5e+18 \\
   3 & 6 & 13 & 137/3 & 6 & 7 & $(H)^{4}\,\bar{\newQ}_R\,q_L$ & 1.6e+18 \\
   6 & 2 & -13 & 178/15 & 10 & 7 & $H^\dagger\,(\bar{d}_R^c\,d_R)\,\bar{\newQ}_L\,e_R$ & 6.9e+18 \\
   $\overline{6}$ & 2 & -17 & 298/15 & 10 & 7 & $H\,(\bar{u}_R^c\,u_R)\,\bar{e}_R\,\newQ_L$ & 1.2e+18 \\
   6 & 3 & -10 & 124/15 & 15 & 7 & $H^\dagger\,(\bar{d}_R^c\,d_R)\,\bar{\newQ}_R\,\ell_L$ & 1.2e+19 \\
   $\overline{6}$ & 3 & -14 & 44/3 & 15 & 7 & $H\,(\bar{u}_R^c\,u_R)\,\bar{\ell}_L\,\newQ_R$ & 1.2e+18 \\
   6 & 4 & -7 & 94/15 & 20 & 7 & $H^\dagger\,(q_L^c\,\bar{q}_L)\,\bar{\newQ}_L\,d_R$ & 8.3e+19 \\
   6 & 4 & 11 & 166/15 & 20 & 7 & $H\,\bar{\ell}_L\,u_R\,\bar{\newQ}_R\,q_L$ & 2.1e+18 \\
   6 & 5 & -4 & 88/15 & 25 & 7 & $H^\dagger\,(\bar{q}_L^c\,q_L)\,\bar{\newQ}_R\,\ell_L$ & 2.4e+18 \\
   $\overline{6}$ & 5 & -2 & 76/15 & 25 & 7 & $H^\dagger\,(\bar{q}_L^c\,q_L)\,\bar{\ell}_L\,\newQ_R$ & 2.4e+18 \\
   $\overline{6}$ & 5 & -8 & 136/15 & 25 & 7 & $H\,(\bar{q}_L^c\,q_L)\,\bar{\ell}_L\,\newQ_R$ & 2.4e+18 \\
   8 & 2 & -15 & 52/3 & 12 & 7 & $H\,\bar{d}_R\,\gamma_\mu\,u_R\,\bar{e}_R\,\gamma^\mu\,\newQ_R$ & 1.1e+18 \\
   8 & 3 & -12 & 112/9 & 18 & 7 & $H\,\bar{\ell}_L\,u_R\,\bar{d}_R\,\newQ_L$ & 1.3e+18 \\
   8 & 4 & -9 & 28/3 & 24 & 7 & $H\,\bar{q}_L\,u_R\,\bar{\ell}_L\,\newQ_R$ & 3.0e+18 \\
   8 & 5 & 0 & 16/3 & 30 & 7 & $H^\dagger\,\bar{q}_L\,\gamma_\mu\,q_L\,\bar{\ell}_L\,\gamma^\mu\,\newQ_L$ & 1.1e+18 \\
   8 & 5 & -6 & 8 & 30 & 7 & $H\,\bar{q}_L\,\gamma_\mu\,q_L\,\bar{\ell}_L\,\gamma^\mu\,\newQ_L$ & 1.1e+18 \\
   10 & 1 & -6 & 4/3 & 15 & 7 & $\bar{\newQ}_L\,e_R\,G^{2}$ & 6.8e+24 \\
   10 & 1 & 12 & 16/3 & 15 & 7 & $H\,\bar{\newQ}_R\,q_L\,(\bar{u}_R^c\,u_R)$ & 3.2e+19 \\
   10 & 3 & -6 & 20/9 & 45 & 7 & $H^\dagger\,\bar{\newQ}_R\,q_L\,(\bar{d}_R^c\,d_R)$ & 1.8e+20 \\
   10 & 4 & -3 & 2 & 60 & 7 & $H^\dagger\,(\bar{q}_L^c\,q_L)\,\bar{\newQ}_L\,d_R$ & 4.9e+18 \\
   10 & 4 & 3 & 2 & 60 & 7 & $H^\dagger\,(\bar{q}_L^c\,q_L)\,\bar{\newQ}_L\,u_R$ & 4.9e+18 \\
   10 & 4 & 9 & 14/3 & 60 & 7 & $H\,(\bar{q}_L^c\,q_L)\,\bar{\newQ}_L\,u_R$ & 1.5e+18 \\
   $\overline{15}$ & 2 & 11 & 65/12 & 40 & 7 & $H^\dagger\,(\bar{d}_R^c\,d_R)\,\bar{u}_R\,\newQ_L$ & 1.2e+18 \\
   $\overline{15}$ & 3 & 8 & 11/3 & 60 & 7 & $H^\dagger\,(\bar{d}_R^c\,d_R)\,\bar{q}_L\,\newQ_R$ & 2.2e+18 \\
   15 & 4 & -5 & 35/12 & 80 & 7 & $H^\dagger\,\bar{\newQ}_R\,q_L\,\bar{q}_L\,d_R$ & 1.3e+18 \\
   15 & 4 & 7 & 47/12 & 80 & 7 & $H\,\bar{\newQ}_R\,q_L\,\bar{q}_L\,u_R$ & 1.3e+18 \\
   15$^\prime$ & 1 & -2 & 2/21 & 35 & 7 & $\bar{\newQ}_L\,d_R\,G^{2}$ & 1.7e+22 \\
   15$^\prime$ & 1 & 4 & 8/21 & 35 & 7 & $\bar{\newQ}_L\,u_R\,G^{2}$ & 4.0e+18 \\
   15$^\prime$ & 2 & 1 & 5/21 & 70 & 7 & $\bar{\newQ}_R\,q_L\,G^{2}$ & 1.3e+18 \\
   24 & 1 & -2 & 8/75 & 50 & 7 & $\bar{\newQ}_L\,d_R\,G^{2}$ & 7.5e+20 \\
   24 & 1 & 4 & 32/75 & 50 & 7 & $\bar{\newQ}_L\,u_R\,G^{2}$ & 7.3e+20 \\
   24 & 2 & 1 & 4/15 & 100 & 7 & $\bar{\newQ}_R\,q_L\,G^{2}$ & 5.5e+18 \\
   27 & 1 & -6 & 1 & 54 & 7 & $\bar{e}_R\,\newQ_L\,G^{2}$ & 2.9e+20 \\
   27 & 2 & -3 & 1/2 & 108 & 7 & $\bar{\ell}_L\,\newQ_R\,G^{2}$ & 4.0e+18 \\
   42 & 1 & -2 & 4/51 & 119 & 7 & $\bar{\newQ}_L\,d_R\,G^{2}$ & 2.5e+18 \\
   42 & 1 & 4 & 16/51 & 119 & 7 & $\bar{\newQ}_L\,u_R\,G^{2}$ & 2.5e+18 \\
   \midrule
   3 & 3 & -20 & 212/3 & 3 & 8 & $(H^\dagger)^{2}\,\bar{\newQ}_L\,d_R\,(\bar{e}_R^c\,e_R)$ & 1.0e+18 \\
   3 & 4 & -17 & 167/3 & 4 & 8 & $(H^\dagger)^{2}\,(\bar{u}_R\,u_R^c)\,\bar{\newQ}_R\,\ell_L$ & 1.2e+18 \\
   $\overline{6}$ & 1 & 16 & 256/15 & 5 & 8 & $(\bar{e}_R^c\,e_R)\,\sigma \cdot G\,\bar{u}_R\,\newQ_L$ & 2.1e+19 \\
   6 & 5 & -10 & 172/15 & 25 & 8 & $(H^\dagger)^{2}\,(\bar{q}_L\,q_L^c)\,\bar{\newQ}_L\,d_R$ & 1.7e+18 \\
   10 & 1 & -12 & 16/3 & 15 & 8 & $\bar{u}_R\,\gamma_\mu\,d_R\,\sigma \cdot G\,\bar{\newQ}_R\,\gamma^\mu\,e_R$ & 3.2e+19 \\
   10 & 2 & -9 & 10/3 & 30 & 8 & $\bar{u}_R\,d_R\,\sigma \cdot G\,\bar{\newQ}_L\,\ell_L$ & 2.0e+19 \\
   10 & 4 & -9 & 14/3 & 60 & 8 & $(H^\dagger)^{2}\,\bar{\newQ}_R\,q_L\,(\bar{d}_R^c\,d_R)$ & 1.5e+18 \\
   15 & 1 & -14 & 49/6 & 20 & 8 & $\bar{\newQ}_Ld_R\,(\bar{e}_R^c\,e_R)\,\sigma \cdot G$ & 1.9e+18 \\
   $\overline{24}$ & 1 & 8 & 128/75 & 50 & 8 & $\bar{u}_R\,\gamma_\mu\,e_R\,\sigma \cdot G\,\bar{d}_R\,\gamma^\mu\,\newQ_R$ & 2.4e+19 \\
   $\overline{24}$ & 1 & -10 & 8/3 & 50 & 8 & $(\bar{d}_R\,d_R^c)\,\sigma \cdot G\,\bar{e}_R\,\newQ_L$ & 3.7e+18 \\
   $\overline{24}$ & 2 & 5 & 68/75 & 100 & 8 & $\bar{d}_R\,\ell_L\,\sigma \cdot G\,\bar{u}_R\,\newQ_L$ & 5.4e+18 \\
   $\overline{24}$ & 2 & -7 & 116/75 & 100 & 8 & $(\bar{d}_R\,d_R^c)\,\sigma \cdot G\,\bar{\ell}_L\,\newQ_R$ & 3.8e+18 \\
   $\overline{24}$ & 3 & 2 & 56/75 & 150 & 8 & $\bar{q}_L\,\ell_L\,\sigma \cdot G\,\bar{d}_R\,\newQ_R$ & 1.3e+18 \\
   24& 3 & 4 & 16/15 & 150 & 8 & $H\,G^2\,\bar{\newQ}_R\,q_L$ & 1.3e+18 \\
   27 & 1 & 0 & 0 & 54 & 8 & $(\bar{d}_R\,d_R^c)\,\sigma \cdot G\,\bar{u}_R\,\newQ_L$ & 3.5e+20 \\
   27 & 3 & 0 & 2/3 & 162 & 8 & $\bar{q}_L\,d_R\,\sigma \cdot G\,\bar{\ell}_L\,\newQ_R$ & 1.1e+18 \\
   27 & 3 & -6 & 5/3 & 162 & 8 & $\bar{q}_L\,u_R\,\sigma \cdot G\,\bar{\ell}_L\,\newQ_R$ & 1.0e+18 \\
   35 & 1 & 0 & 0 & 105 & 8 & $\bar{\newQ}_L\,u_R\,(\bar{d}_R^c\,d_R)\,\sigma \cdot G$ & 3.8e+18 \\
   35 & 1 & 6 & 2/3 & 105 & 8 & $(\bar{u}_R^c\,u_R)\,\bar{\newQ}_L\,d_R\,\sigma \cdot G$ & 3.8e+18 \\
   $\overline{42}$ & 1 & 8 & 64/51 & 119 & 8 & $(\bar{d}_R^c\,d_R)\,\sigma \cdot G\,\bar{u}_R\,\newQ_L$ & 2.1e+18 \\
   \midrule
   $\overline{3}$ & 1 & 20 & 200/3 & 1 & 9 & $(\bar{d}_R^c\,d_R)\,(\bar{e}_R^c\,e_R)\,\bar{u}_R\,\newQ_L$ & 6.2e+19 \\
   3 & 1 & 22 & 242/3 & 1 & 9 & $\bar{\newQ}_L\,u_R\,(\bar{\ell}_L\,\ell_L^c)\,(\bar{e}_R\,e_R^c)$ & 2.0e+19 \\
   3 & 2 & -23 & 269/3 & 2 & 9 & $(\bar{u}_R\,u_R^c)\,(\bar{e}_R^c\,e_R)\,\bar{\newQ}_R\,\ell_L$ & 1.3e+18 \\
   6 & 1 & -22 & 484/15 & 5 & 9 & $(\bar{u}_R\,u_R^c)^{2}\,\bar{\newQ}_L\,e_R$ & 1.6e+18 \\
   $\overline{6}$ & 1 & -20 & 80/3 & 5 & 9 & $(\bar{u}_R^c\,u_R)\,(\bar{d}_R\,\gamma_\mu\,u_R)\,(\bar{e}_R\,\gamma^\mu\,\newQ_R)$ & 3.0e+18 \\
   8 & 1 & -18 & 24 & 6 & 9 & $(\bar{e}_R\,e_R^c)^{2}\,\sigma \cdot G\,\bar{e}_R\,\slashed{\partial}\,\newQ_R$ & 2.5e+18 \\
   10 & 1 & -18 & 12 & 15 & 9 & $(\bar{u}_R\,u_R^c)^{2}\,\bar{\newQ}_L\,d_R$ & 1.3e+18 \\
   $\overline{10}$ & 1 & -18 & 12 & 15 & 9 & $(\bar{u}_R^c\,u_R)\,(\bar{d}_R\,d_R^c)\,\bar{e}_R\,\newQ_L$ & 1.3e+18 \\
   $\overline{21}$ & 1 & 8 & 16/15 & 70 & 9 & $(q_L^c\,\bar{q}_L)^{2}\,\bar{u}_R\,\newQ_L$ & 2.6e+19 \\
   21 & 1 & -2 & 1/15 & 70 & 9 & $\bar{\newQ}_L\,d_R\,G^{3}$ & 2.8e+19 \\
   21 & 1 & 4 & 4/15 & 70 & 9 & $\bar{\newQ}_L\,u_R\,G^{3}$ & 2.8e+19 \\
   $\overline{21}$ & 2 & 5 & 17/30 & 140 & 9 & $(\bar{u}_R\,u_R^c)\,(\bar{d}_R\,d_R^c)\,\bar{q}_L\,{\newQ}_R$ & 1.4e+18 \\
   $\overline{21}$ & 2 & -1 & 1/6 & 140 & 9 & $(\bar{q}_L\,q_L^c)\,(\bar{d}_R\,d_R^c)\,\bar{q}_L\,\newQ_R$ & 1.4e+18 \\
   35 & 1 & -6 & 2/3 & 105 & 9 & $\bar{\newQ}_L\,e_R\,G^{3}$ & 3.8e+18 \\
\end{longtable}
\end{center}

\bibliographystyle{JHEP_mod}
\bibliography{auto,custom}

\end{document}

%% file: macros.tex
\newcommand{\emth}[1]{\ensuremath{#1}\xspace}
\newcommand{\mtx}[2]{#1_\text{#2}}
\newcommand{\newmm}[2]{\newcommand{#1}{\emth{#2}}}
\newcommand{\code}[1]{\texttt{#1}\xspace}
\DeclareRobustCommand{\cpp}{C\nolinebreak[4]\hspace{-0.05em}\raisebox{0.4ex}{\tiny\textbf{++}}\xspace}
\newcommand{\orcid}[1]{\xspace\href{https://orcid.org/#1}{\color[HTML]{A6CE39}\faOrcid{}}\xspace}
\newcommand{\deco}{\code{DECO}}
\newcommand{\form}{\code{FORM}}
\newcommand{\mimes}{\code{MiMeS}}
\definecolor{jubilee}{RGB}{134,143,152}
\newcommand{\ccol}{\color{jubilee}}

\newmm{\mpl}{\mtx{M}{Pl}}
\newmm{\MP}{\mtx{M}{P}}
\newmm{\mP}{\mtx{m}{P}}

\newmm{\lagr}{\mathcal{L}}
\newmm{\qEM}{\mtx{q}{EM}}
\newmm{\alphaEM}{\mtx{\alpha}{EM}}
\newmm{\alphaS}{\mtx{\alpha}{s}}
\newmm{\LLP}{\mtx{\Lambda}{LP}}
\newmm{\LPthresh}{\mtx{\Lambda}{thr}}
\newmm{\Ldec}{\mtx{\Lambda}{EFT}}
\newmm{\Tdec}{\mtx{T}{dec}}

\newmm{\newQ}{\mathcal{Q}}
\newcommand{\newQs}{{\newQ}s\xspace}
\newmm{\NQ}{N_\newQ}
\newmm{\mQ}{m_\newQ}
\newcommand{\gr}[1]{\mathrm{#1}}
\newmm{\dynk}{T}
\newmm{\UPQ}{\mtx{\gr{U}(1)}{PQ}}
\newmm{\NDW}{\mtx{N}{DW}}
\newmm{\dmin}{\mtx{d}{min}}
\newmm{\muH}{\mtx{\mu}{H}}

\newcommand{\ax}{a}
\newmm{\fax}{f_\ax}
\newmm{\ma}{m_\ax}
\newmm{\cagg}{C_{\ax\gamma}}
\newmm{\caggzero}{C_{\ax\gamma,0}^\text{NLO}}
\newmm{\gagg}{g_{\ax\gamma}}
\newmm{\gaee}{g_{\ax{}e}}
\newmm{\pag}{\prob_{\gamma \leftrightarrow a}}
\newmm{\afrac}{\eta_a}
\newmm{\thetai}{\theta_0}
\newmm{\thetaieff}{\mtx{\theta}{eff}}

\newmm{\sigv}{\langle\sigma v\rangle}
\newmm{\gRho}{g_\rho}
\newmm{\gS}{g_s}
\newmm{\rhoR}{\mtx{\rho}{R}}
\newmm{\rhoSM}{\mtx{\rho}{SM}}
\newmm{\Neff}{\mtx{N}{eff}}
\newmm{\TBBN}{\mtx{T}{BBN}}
\newmm{\OmegaDM}{\mtx{\Omega}{c}h^2}

\newcommand{\dd}{\mathrm{d}}
\newcommand{\ee}{\mathrm{e}}
\newcommand{\ii}{\mathrm{i}}

\newcommand{\pref}{``preferred''\xspace}